\documentclass[secnumarabic,twocolumn, graphics,floatfix,tightenlines,aps,prb]{revtex4-1}

 \pdfoutput=1

\usepackage{epsfig}
\usepackage{amsmath,amssymb,amsfonts}
\usepackage{hyperref}
\usepackage{mathrsfs}
\usepackage{bbm}
\usepackage{slashed}
\usepackage{graphicx}
\usepackage{verbatim}
\usepackage[usenames]{color}

\newcommand{\scr}[1]{\mathcal{#1}}

\newcommand{\LCm}{{\scriptscriptstyle -}} %LC supersripts
\newcommand{\LCp}{{\scriptscriptstyle +}}

\newcommand{\LCpm}{{\scriptscriptstyle \pm}}

\newcommand{\bra}[1]{\langle #1|}
\newcommand{\ket}[1]{|#1\rangle}

\newcommand{\ve}[1]{{\bf{#1}}}

\newcommand{\av}[1]{\langle #1 \rangle}

\newcommand{\Q}{\pmb{\mathcal{Q}}}
\newcommand{\sQ}{\mathcal{Q}}

\begin{document}
	\title{Effective model for a supercurrent in a pair-density wave}
	\author{Jonatan W\aa rdh}
	\email[]{jonatan.wardh@physics.gu.se}
	\affiliation{Department of Physics, University of Gothenburg,
		SE-41296 Gothenburg, Sweden}
	\author{Mats Granath}
	\email[]{mats.granath@physics.gu.se}
	\affiliation{Department of Physics, University of Gothenburg,
		SE-41296 Gothenburg, Sweden}
	
\begin{abstract}
We extend the standard effective model of d-wave superconductivity of a single band tight-binding Hamiltonian with a nearest-neighbor attraction to include finite range periodically modulated pair-hopping. The pair-hopping is characterized by a fixed wave number $\Q=\sQ\hat{x}$ breaking lattice rotational symmetry. Within self-consistent BCS theory we study the general variational state consisting of two incommensurate singlet pair-amplitudes $\Delta_{\ve{Q}_1}$ and $\Delta_{\ve{Q}_2}$ and find two types of ground states; one of the Larkin-Ovchnnikov (LO) or pair-density wave (PDW) type with $\Delta_{\ve{Q}_1}=\Delta_{\ve{Q}_2}$ and $\ve{Q}_1=-\ve{Q}_2\approx \Q$, and one of the Fulde-Ferrell (FF) type with $\Delta_{\ve{Q}_2}=0$ and $\ve{Q}_1\approx \pm \Q$. An anomalous term in the static current operator arising from the pair-hopping ensures that Bloch's theorem on ground state current is enforced also for the time-reversal and parity breaking FF state, despite no spin-population imbalance. We also consider a supercurrent by exploring the space of pair-momenta $\ve{Q}_1$ and $\ve{Q}_2$ and identify characteristics of a state with multiple finite momentum order-parameters. This includes the possibility of phase-separation of current densities and spontaneous mirror-symmetry breaking manifested in the directional dependence of the depairing current. 
\end{abstract}
\pacs{}
\maketitle
%

% Convetions. Phase-separation, d-wave  Larkin-Ovchinnikov

\section{Introduction \label{intro}}
It is well established that various coexisting orders, in particular, spin- and charge-density wave order, are a ubiquitous phenomenon of the cuprate high-temperature superconductivity. Less clear is the degree of interdependency between these orders and superconductivity.\cite{fradkin2015colloquium,tranquada2004quantum,mesaros2016commensurate,tranquada1995evidence,da2014ubiquitous,comin2015symmetry} Regardless of the detailed microscopic physics, for a state with coexisting superconductivity (SC) and charge-density wave (CDW), the superconducting order must be modulated with a corresponding wave length, which has indeed been observed recently.\cite{hamidian2016detection} A distinct state where the superconducting order is modulated around a mean of zero has also been discussed, referred to as a pair-density wave (PDW) state.\cite{himeda2002stripe,berg2009theory,berg2007dynamical} This state is suggested to play a significant role for the anomalous suppression of superconductivity in LBCO at $1/8$ doping\cite{moodenbaugh1988superconducting,tranquada1995evidence} by decoupling the CuO$_2$ layers.\cite{li2007two,berg2007dynamical}

The PDW order is a unidirectional singlet superconducting order that varies in space as $\Delta(\ve{r})=\Delta_{\ve{Q}}\cos(\ve{Q} \cdot \ve{r})$  (with $\ve{r}$ as the center of mass coordinate). For superconductors with Zeeman split population of spins, finite momentum pairing is a natural consequence of mismatched time-reversed Fermi-surfaces.\cite{fulde1964superconductivity,larkin1965inhomogeneous} Although in cuprates there is no symmetry breaking field, and the physics may be quite different, similar states are discussed. Here one may distinguish between two types of states. The Larkin-Ovchinnikov (LO) state,\cite{larkin1965inhomogeneous} with two pair-fields  $\Delta_{\ve{Q}}=\Delta_{-\ve{Q}}$ and broken translational symmetry, is the PDW defined above. The Fulde-Ferrell (FF) state,\cite{fulde1964superconductivity} with one pair-field $\Delta_{-\ve{Q}}=0$ such that $\Delta(\ve{r})=\Delta_{\ve{Q}}e^{i \ve{Q} \cdot \ve{r}}$, is translational invariant  but breaks time-reversal and parity. From a symmetry perspective FF is identical to a current-carrying SC state.

A PDW without coexisting uniform order has not been directly observed, and neither is a mechanism for its formation in systems without population imbalance clear.\cite{fradkin2015colloquium} Such a state was first suggested in variational Monte Carlo study of the 2D t-t'-J model.\cite{himeda2002stripe} From variational calculations using tensor networks it is also clear that a striped PDW is near degenerate with the uniform d-wave superconductor.\cite{corboz2014competing} On the other hand, from DMRG studies of t-J ladders the evidence for PDW order is weaker.\footnote{For a recent study see \citet{dodaro2016intertwined}} Regardless of the microscopic mechanism, the implications of a PDW state has been explored in some detail. In 2D the PDW becomes unstable to topological excitations and a rich phase diagram develops as the PDW melts.\cite{berg2009charge,barci2011role,agterberg2008dislocations} The natural Fermi arc type Fermi-surface of the PDW has also been discussed,\cite{baruch2008spectral} as has the possible connection to quantum oscillations in high magnetic fields.\cite{zelli2012quantum} In addition, there are suggestions of a close connection between a PDW state and the elusive pseudogap state, possibly of a form with broken time-reversal symmetry.\cite{lee2014amperean,agterberg2015emergent,wang2015coexistence,wang2015interplay}

The aim of this paper is to present an interacting Hamiltonian with a PDW BCS-like ground state and study its destruction as a function of current. In ordinary BCS-theory with an attractive interaction there is an infinite susceptibility towards forming Cooper-pairs with zero momentum, $\av{c_{\ve{k}} c_{-\ve{k}}}$, due to perfect nesting of the Fermi-surface. For finite momentum, $\av{c_{\downarrow \ve{k}+\ve{Q}/2}c_{\uparrow -\ve{k}+\ve{Q}/2}}$, this is in general not the case, i.e. a weak coupling instability would require fine tuning and we have to consider finite interaction strength.\footnote{Ordinary BCS-theory is valid in the weak coupling limit. However, even at finite coupling the mean-field calculation is known to give qualitatively right results for zero-momentum pairing, at least for temperatures $T\ll T_c$ where the influence of collective excitations can be considered low\cite{Engelbrecht1997}}

Even with finite interaction strength, it is unexpected that $\ve{Q}  \neq 0$ would be preferable over $\ve{Q}=0$ since portions of the Fermi-surface will remain ungapped. A possible exception could be for an interaction that promotes d-wave order; the already reduced nodal gap might conspire with a finite momenta condensate and form an effectively less gapped Fermi-surface. Indeed such a result was reported by \citet{loder2010} who find that for sufficiently strong nearest-neighbor attraction finite momentum pairing triumphs over zero-momentum. However, we find the interaction strength needed to make finite momentum pairing a global minimum to be higher than reported (see Appendix \ref{nearest_finite_q} for details).

In order to consider moderate interaction strengths we tailor a Hamiltonian which promotes PDW order. As inspiration, we take stripe domain walls acting as $\pi$-junctions of superconducting order.\cite{berg2009theory} The interaction is a generalized nearest-neighbor attraction which includes periodically modulated pair-hopping with period and range similar to experimentally observed stripe periods.\cite{tranquada2007neutron} The model breaks lattice rotational invariance by assuming that pair-hopping is modulated (with wavenumber $\Q$) along one of the crystal axes, but preserves translational invariance. This would be consistent with a preexisting nematic order with susceptibility to smectic (PDW or stripe) order of a certain wavelength.\cite{fradkin2010nematic} The model allows for doing self-consistent calculations of a PDW state and explore the doping and other parameter dependence. Unexpectedly, we find that the lowest energy PDW/LO state is typically near degenerate with an FF state, and a phase transition between the two states may occur as a function of doping (see FIG \ref{phase}). The FF state carries zero current (despite breaking time-reversal and parity symmetry) due to an anomalous current arising from the interaction, consistent with Bloch's theorem on the absence of ground state current.\cite{bohm1949note,ohashi1996bloch}  
\begin{figure}[h!]
	\centering
	\includegraphics[width=0.45\textwidth]{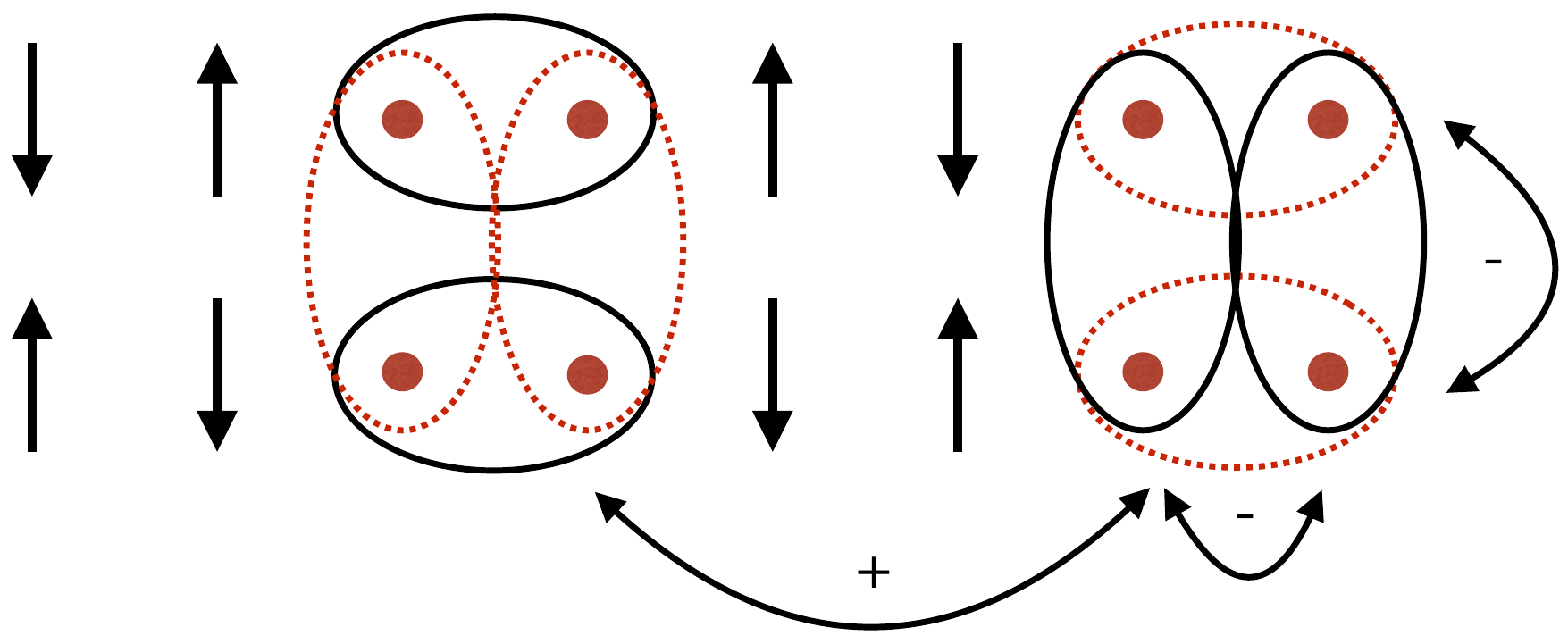}
	\caption{\label{hopping_fig}(Color online) Sketch of the interaction \eqref{pair-hopping} with local attraction and longer range phase-flip pair-hopping. Solid (dashed) rings indicate positive (negative) pair-amplitudes of a commensurate pair-density wave. }
\end{figure}

Having access to an interaction it is possible to generalize the standard formalism for a uniform current in a superconductor by allowing both the momenta and magnitudes of the two coupled order-parameters of the LO state to be varied self-consistently. From this, we identify two characteristic features that may be found in a multi-component finite momentum pairing superconductor. Considering current $J_x$ along $\Q = \sQ \hat{x}$ we find, for a certain parameter regime, a first order phase transition as function of current between an LO and FF state  (see FIG \ref{J_x_energy} and \ref{J_x_order}). Consequently, for a range of currents, we anticipate a phase-separated state with inhomogeneous current density. For a different parameter regime, and for currents close to the deparing current, we find two distinct LO states that are related by mirror-symmetry with respect to $\hat{x}$. Consequently, there is a cusp in the directional dependence of depairing current (see FIG \ref{all_angles} and \ref{cusp}) as the system switches between these two branches. In principle, this may also manifest itself in terms of a spontaneous transverse current ($J_y$). 

This paper is organized as follows. In Section \ref{inter} the proposed Hamiltonian is discussed and its mean-field decomposition is introduced in \ref{mean-field}, where approximations are discussed. The ground state and phase diagram of the model is presented in \ref{ground_state} and the cancellation of current in the FF state is discussed in relation to Bloch's theorem on ground state current in Section \ref{current_bloch}. We proceed in Section \ref{current_section} to discuss the generalized description of a current carrying state with two order-parameters, the results are presented in \ref{phase_sep} and \ref{current_all_angles}. We conclude with a summary and an outlook in Section \ref{discussion}.

\section{Model Hamiltonian\label{inter}}

We start with a tight-binding Hamiltonian on a square lattice (length $a=1$)
\begin{equation}
H =  \sum_{\ve{k} \sigma} \varepsilon(\ve{k}) c^\dagger_{\ve{k} \sigma} c_{\ve{k} \sigma} +H_{\text{int}}
\label{Ham}
\end{equation}
with the dispersion
$\varepsilon(\ve{k})= -2t (\cos(k_x)+\cos(k_y) )-4 t' \cos(k_x) \cos(k_y)$ where $t'=-0.3 t$ and $t=1$. (All energies will be measured in units of $t$.) The interaction is given by 
 \begin{equation}
\begin{split}
&H_{\text{int}} =\\
& - V \sum_{ijkl} \sum_{\sigma, \sigma'}T(\ve{r}_{ij}^{\LCp} \LCm \ve{r}_{kl}^{\LCp})t(\ve{r}_{ij}^{\LCm}, \! \ve{r}_{kl}^{\LCm}) c^{\dagger}_{\sigma, i}c^{\dagger}_{\sigma' \!, j}c_{\sigma' \!,l }c_{\sigma,k}  \, ,
\end{split}
\label{interaction}
\end{equation}
with $\ve{r}_{ij}^{\LCpm}=(\ve{r}_i \pm \ve{r}_j)/2$, illustrated in FIG \ref{hopping_fig}.
 \footnote{The arrows illustrate a possible coexisting striped magnetic order, which is not considered in the present work.}
  $\ve{r}_{ij}^{\LCm}$ refers to the relative coordinate between electrons with $t(\ve{r}_{ij}^{\LCm}, \! \ve{r}_{kl}^{\LCm})  = \delta_{\ve{r}_{ij}^{\LCm},\hat{x}/2} \delta_{\ve{r}_{kl}^{\LCm},\hat{x}/2} + (\hat{x} \rightarrow - \hat{x}, \pm \hat{y})$ assuring pair-creation and annihilation on nearest-neighboring sites. The pair-hopping is accounted for by $T(\ve{r}_{ij}^{\LCp} \LCm \ve{r}_{kl}^{\LCp})$ where $\ve{r}_{ij}^{\LCp} \LCm \ve{r}_{kl}^{\LCp}$ is the relative coordinate between pairs. This defines a class of Hamiltonians without explicitly broken translational invariance. To make contact with observed striped orders we consider an explicitly broken rotational invariance with pair-hopping in the $x$-direction with length $P/2$,\footnote{Here we interpret the argument in $T$ as mapped to the set $x \in (-L/2 , L/2)$ where $L$ is the length of the crystal.}
\begin{equation}
\begin{split}
&T (\ve{r}_{ij}^{\LCp} \LCm \ve{r}_{kl}^{\LCp})= \\
&\frac{\kappa_y \kappa_x}{2\pi} e^{- \sum \limits_{\mu}\kappa_\mu^2 \frac{(r_{ij,\mu}^{\LCp} \LCm r_{kl,\mu}^{\LCp})^2}{2} }
\cos \sQ \! (r_{ij,x}^{\LCp} \LCm r_{kl,x}^{\LCp}) \, ,
\end{split}
\label{pair-hopping}
\end{equation}
where $\Q=\sQ \hat{x}=\frac{2 \pi}{P} \hat{x}$ represents the modulation ( $P=8$ is considered in this paper). Here $\kappa_\mu, \mu = x,y$ sets the hopping range and in the limit $\kappa_{x,y} \rightarrow \infty$ \eqref{pair-hopping} reduces to an ordinary nearest-neighbor interaction. 
 In order to have a negligible zero-momentum pairing the modulation needs to be well resolved by the hopping range, thus we will consider  $\kappa_x \lesssim \frac{2}{P}$.
%
%\com{ In order to resolve the modulation the hopping range needs to be roughly one period or greater, thus we will consider  $\kappa_x \lesssim \frac{2}{P}$}.
%The hopping range along $x$ should be at least as long as one modulation period, thus  $\kappa_x \lesssim \frac{2}{P}$.
 We have also included a possible finite hopping range along the $y$-direction, but it is of secondary importance and the value $\kappa_y$ will only be specified when it is essential.

Going over to reciprocal space and identifying the singlet paring we find
\begin{equation}
\begin{split}
&H_{\text{int}}= \\
&  \frac{1}{N}\sum_{\ve{k}, \ve{k}' ,\ve{q}}  V(\ve{k},\ve{k}',\ve{q}) c^{\dagger}_{\uparrow, \ve{k}+\frac{\ve{q}}{2}}c^{\dagger}_{\downarrow, -\ve{k}+\frac{\ve{q}}{2}}c_{\downarrow, -\ve{k}'+\frac{\ve{q}}{2}}c_{\uparrow, \ve{k}'+\frac{\ve{q}}{2}} \\
\end{split} \, ,
\label{Ham2}
\end{equation}
where
\begin{equation}
V(\ve{k},\ve{k}',\ve{q})= -V v(\ve{q}) 
\left(g_d(\ve{k}) g_d(\ve{k}') +g_s(\ve{k}) g_s(\ve{k}') \right)\label{V_form} \, .
\end{equation}
Here $g_d(\ve{k})=\cos(k_x)-\cos(k_y), g_s(\ve{k})=\cos(k_x)+\cos(k_y)$ and
\begin{equation}
v(\ve{q}) =e^{-\frac{q_y^2}{2\kappa_y^2}}\left(e^{-\frac{(q_x-\sQ)^2}{2 \kappa_x^2}}+e^{-\frac{(q_x+\sQ)^2}{2 \kappa_x^2}} \right) 
\label{gauss-potential}
\end{equation}
is a Gaussian potential that benefits finite momentum pairing, with $\kappa_{x,y}$ acting as the potential-width. Subsequently, we will only present and discuss the $d$-wave part explicitly since the $s$-wave turns out to be negligible.

\subsection{\label{mean-field}Mean-field solution}
We will study the interacting model in BCS mean-field theory using the following quadratic Hamiltonian
\begin{equation}
\begin{split}
&H_{\text{MF}}= \\
&\sum_{\ve{k}} (\varepsilon_{\ve{k}}-\mu)n_{\ve{k}} \! + \! \! \! \! \! \! \! \!\sum \limits_{\ve{q}=\ve{Q}_1,\ve{Q}_2}  \! \!\! \!\! \!\left(\Delta^*_{\ve{q}}(\ve{k}) c_{\downarrow, -\ve{k}+\frac{\ve{q}}{2}}c_{\uparrow ,\ve{k}+\frac{\ve{q}}{2}}+\text{h.c.} \right) ,
\label{eff_H}
\end{split}
\end{equation}
with $n_{\ve{k}}=\sum_{\sigma}c^\dagger_{\ve{k} \sigma} c_{\ve{k} \sigma}$ where $\mu$ is tuned to get the correct particle number (see Appendix \ref{appendix_mean_field}). That there are two pair-fields with different momenta $\ve{Q}_1$ and $\ve{Q}_2$ complicates the model compared to standard BCS, as CDW operators $\rho_{\ve{Q}_{\text{CDW}}}=\sum_{\sigma}c^\dagger_{\ve{k}\sigma}c_{\ve{k}+\ve{Q}_{\text{CDW}}\sigma}$ (with $\ve{Q}_{\text{CDW}}=n(\ve{Q}_1-\ve{Q}_2), \, n \in \mathbb{Z}$ ) and higher order pair-fields of the type $\Delta_{\ve{q}+\ve{Q}_{\text{CDW}}}$ are induced. (For the FF state with only one pair-field, no other terms are generated, and \eqref{eff_H} is complete.)
 Note that other possible self-consistent charge or spin orders (such as the striped magnetic order shown in FIG \ref{hopping_fig}) are not considered.
 In Appendix \ref{appendix_mean_field} we show that the CDW fields are expected to be small, partly due to the particular form of the interaction studied, and will be neglected. Also discussed there, is a nematic distortion of the single-particle energies which we have not included in the present study. 

With these approximations, the self-consistency equation for the gap functions is given by
\begin{equation}
\Delta_{\ve{q}}(\ve{k})=\frac{1}{N}\sum_\ve{k'}V(\ve{k},\ve{k}',\ve{q})\av{c_{\downarrow, -\ve{k}'+\frac{\ve{q}}{2}}c_{\uparrow ,\ve{k}'+\frac{\ve{q}}{2}}} \, .
\label{self_consistent}
\end{equation}

Even if the additional CDW and pair-fields are set to zero, the Hamiltonian \eqref{eff_H} cannot be directly diagonalized except for short commensurate periods. Since we are interested in studying the system for continuous (incommensurate) variations of the pair momenta we have to truncate the matrix form of the Hamiltonian. A convenient way to formalize this is to work with the Gorkov equations for the single-particle Greens functions. The spin-independent imaginary time Greens function takes the form $\scr{G}_{\ve{k},\ve{k}'}(\tau)=-\av{ T_{\tau} c_{ \sigma ,\ve{k}}(\tau) c^\dagger_{\sigma, \ve{k}'}(0)}$ with $\sigma = \,  \downarrow, \uparrow$ and the anomalous Greens functions $\scr{F}_{\ve{k},\ve{k}'}(\tau)=-\av{ T_{\tau} c_{\downarrow ,-\ve{k}}(\tau) c_{\uparrow, \ve{k}'}(0)} $, $\scr{F}^{*}_{\ve{k},\ve{k}'}(\tau)=\av{ T_{\tau} c^\dagger_{\downarrow ,-\ve{k}}(\tau) c^\dagger_{\uparrow, \ve{k}'}(0)} $.

The general expressions for the full Greens functions in terms of Matsubara frequencies $z=i\omega_n$ ($\omega_n=\frac{2\pi}{\beta}(n+\frac{1}{2})$) are derived as
\begin{eqnarray}
\scr{G}_{\ve{k},\ve{k}'}(z) &=& G_{0.\ve{k}}(z) \bigg( \delta_{\ve{k},\ve{k}'}+\sum_{\ve{q}}\Delta_{\ve{q}}(\ve{k}-\frac{\ve{q}}{2})\scr{F}^*_{\ve{k}-\ve{q},\ve{k}'}(z) \bigg) \; \; \; \;  \; \label{off-d}\\
\scr{F}^*_{\ve{k},\ve{k}'}(z)&=&-G_{0.\ve{k}}(-z)\sum_{\ve{q}}\Delta_{\ve{q}}^*(\ve{k}+\frac{\ve{q}}{2})\scr{G}_{\ve{k}+\ve{q},\ve{k}'}(z) \label{off-d2}\, .
\end{eqnarray}
Considering the diagonal part ($\ve{k}=\ve{k}'$) of the full Greens function
\begin{eqnarray}
\scr{G}_{\ve{k}}(z) &=& G_{0.\ve{k}}(z) \bigg( 1-\sum_{\ve{q},\ve{q}'}\Delta_{\ve{q}}(\ve{k}-\frac{\ve{q}}{2}) \Delta_{\ve{q}'}^*(\ve{k}+\frac{\ve{q}}{2}'-\ve{q})  \nonumber\\
  &\times & G_{0.\ve{k}-\ve{q}}(-z)\scr{G}_{\ve{k}+\ve{q}'-\ve{q},\ve{k}}(z)   \bigg)
\end{eqnarray}
we see that it couples to off-diagonal Greens functions, with a static part corresponding to the $\rho_{\ve{Q}_{\text{CDW}}}$. Consistent with the discussion of the mean-field Hamiltonian these off-diagonal Greens functions induce correspondingly shifted anomalous Greens functions. With the same motivation, we will truncate this proliferation at the lowest level neglecting all off-diagonal Greens functions. With this approximation (also used by \citet{loder2010}) we find a simple expression for the diagonal part of the Greens function and a self-consistency equation of the form
\begin{eqnarray}
\label{approx_1}
\scr{G}_{\ve{k}}(z)&=& \! \! \Big( G^{-1}_{0.\ve{k}}(z)+\sum_{\ve{q}}|\Delta_{\ve{q}}(\ve{k}-\frac{\ve{q}}{2})|^2G_{0.\ve{k}-\ve{q}}(-z) \Big)^{-1} \quad \\
\label{approx_2}
\Delta_{\ve{q}}(\ve{k})&=&-\frac{1}{N\beta}\sum_{\ve{k}',z} V(\ve{k},\ve{k}',\ve{q}) \scr{F}_{\ve{k}'-\frac{\ve{q}}{2},\ve{k}'+\frac{\ve{q}}{2}}(z) \, .
\end{eqnarray}

The Matsubara sum is conveniently handled by analytic continuation 
%giving a form 
where we can sum over residues of the anomalous Greens function. We evaluate this system over a $350 \times 350$ grid, at zero and finite temperature.

\subsection{\label{ground_state}Ground state, phase diagram}

To solve the system, \eqref{approx_1} and \eqref{approx_2} was iterated until convergence. Stable solutions was found for $V \gtrsim 1.2$ and $V=1.5$ is used throughout the paper. Assuming $\Delta_{\ve{Q}_1}, \Delta_{\ve{Q}_2}$ with $\ve{Q}_1=-\ve{Q}_2$ two types of locally stable states could be identified: one LO state which breaks translation invariance $\Delta_{\ve{Q}_1} = \Delta_{\ve{Q}_2}$, but also an FF state with $\Delta_{\ve{Q}_1} = 0, \Delta_{\ve{Q}_2} \neq 0$ or $\Delta_{\ve{Q}_2} = 0, \Delta_{\ve{Q}_1} \neq 0$, which breaks time-reversal and parity but preserves translational invariance. The self-consistency relation \eqref{self_consistent} only ensures local stability and $\ve{Q}_1= (Q,0)$ was varied for both LO and FF to find the global minimum in energy
\begin{equation}
E=\frac{1}{N}\sum_{\ve{k} \sigma} \varepsilon(\ve{k}) n_{\ve{k} \sigma} - \! \! \! \sum \limits_{\ve{q}=\ve{Q}_1,\ve{Q}_2 } \frac{|\Delta_{\ve{q}}|^2}{V v(\ve{q})} \,  .
\label{energy}
\end{equation}

\begin{figure}[h!]
	\centering
	\includegraphics[width=0.5\textwidth]{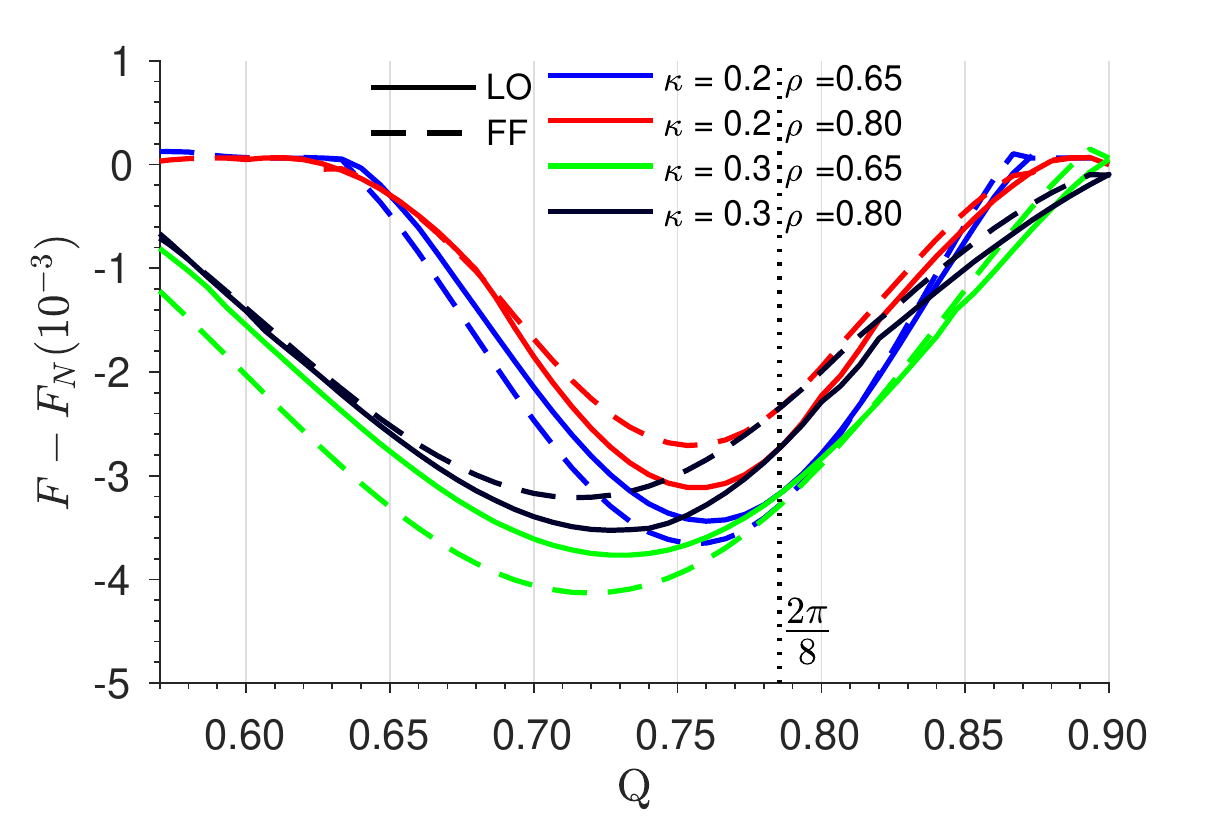}
	\caption{\label{potential} (Color online) Free energy as a function of $\ve{Q}=Q \hat{x}$ for $\rho = 0.8; 0.65, \kappa =0.2 ; 0.3, T=0.01 $. The solid lines correspond to LO with $\Delta_{-\ve{Q}}=\Delta_{\ve{Q}}$ and the dashed lines to FF with $\Delta_{-\ve{Q}}=0$.}
\end{figure}

In FIG \ref{potential} the equilibrium state free energy $F = E - TS$ is plotted against $Q$ for $\kappa_x = 0.2; 0.3$ and electron densities $\rho=0.65 ; 0.80$. A finite temperature $T=0.01$ was used as a regularization (see Appendix \ref{appendix_entropy} for details on the entropy). The lowest energy state $\ve{Q}_0=(Q_{0} ,0)$ coincides well with $\sQ=\frac{2 \pi}{8}$, however there is a slight deviation which grows with the width of the potential, $\kappa_x$. Note that we loose the ability to form a condensate for $|Q -Q_{0}  | \gtrsim \kappa_x$ and there is a negligible zero-momentum pairing. For $\rho = 0.8$ LO is the ground state, but at $\rho = 0.65$ FF has the lowest energy, nevertheless, they are near degenerate. In fact, the LO or FF character of the ground state is not heavily dependent on $\kappa_x$, and we set $\kappa_x= 0.2$ throughout the paper.

\begin{figure}[h!]
	\centering
	\includegraphics[width=0.5\textwidth]{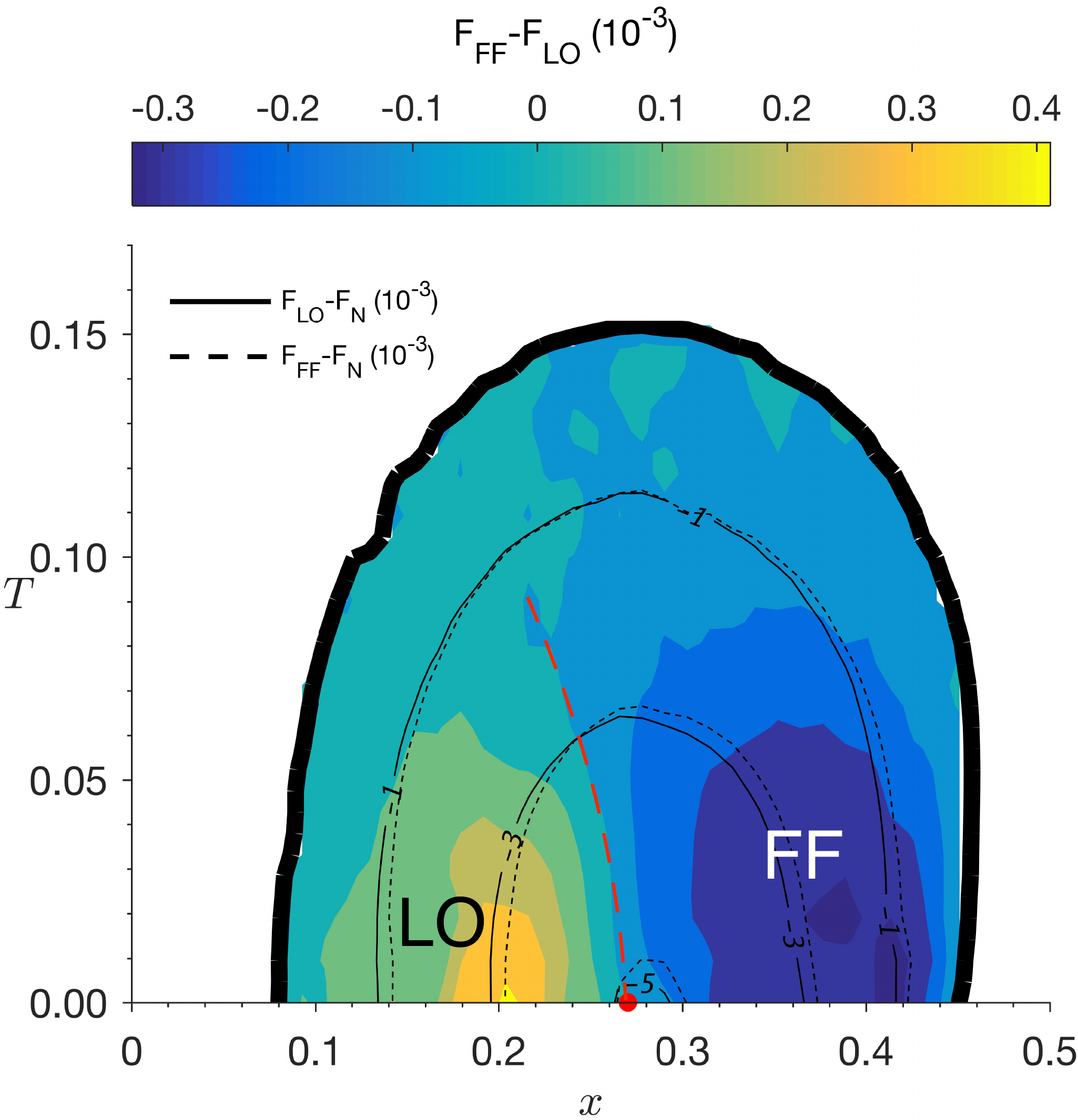}
	\caption{\label{phase}  (Color online) Phase diagram for $\sQ = 2\pi / 8 $, $V = 1.5$, $\kappa_x = 0.2$. The color shading represents the energy difference between LO and FF. The full (dashed) lines represent the energy difference between LO (FF) and normal state respectively. The red dot indicates the quantum critical point and the red dashed line shows the approximate phase-transition which cannot be identified for higher temperature within our numerical precision.}
\end{figure}

In FIG \ref{phase} we report the phase diagram
\footnote{The transition temperatures ($T_c$) presented in FIG 3 are mean-field values ($T_c^{\text{MF}}$), which are expected to be an overestimation of the model's true $T_c$. To better account for the $T_c$ one needs to take phase-fluctuations into consideration.\cite{emery1995importance,berg2009charge,barci2011role,agterberg2008dislocations}} 
for doping $x = 1 - \rho$ and temperature $T$. The free energy $F$ was calculated for LO, FF and the normal state respectively. We see the emergence of two domes where LO and FF dominate respectively, and a region of near degeneracy. At low temperature, we can identify a phase transition between the two distinct symmetry states with a quantum critical point at  $x \sim 0.27$. We have noted that the phase diagram is quite sensitive to the interaction modulation momentum $\Q$, but we will not explore that dependence in present work.

At higher dopings, FF is the ground state despite it breaking time-reversal symmetry. In general, one would expect such a time-reversal breaking state to carry current. However, this is not possible due to the theorem, attributed to Bloch, stating that even with interaction included the ground state must have zero current.\cite{bohm1949note,ohashi1996bloch} For the present model, and for the FF state, this theorem is obeyed due to the appearance of an anomalous current emanating from the interaction which cancels the ordinary current. This will be the topic of Section \ref{current_bloch}.
\begin{figure}[h!]
	\centering
	\includegraphics[width=0.5\textwidth]{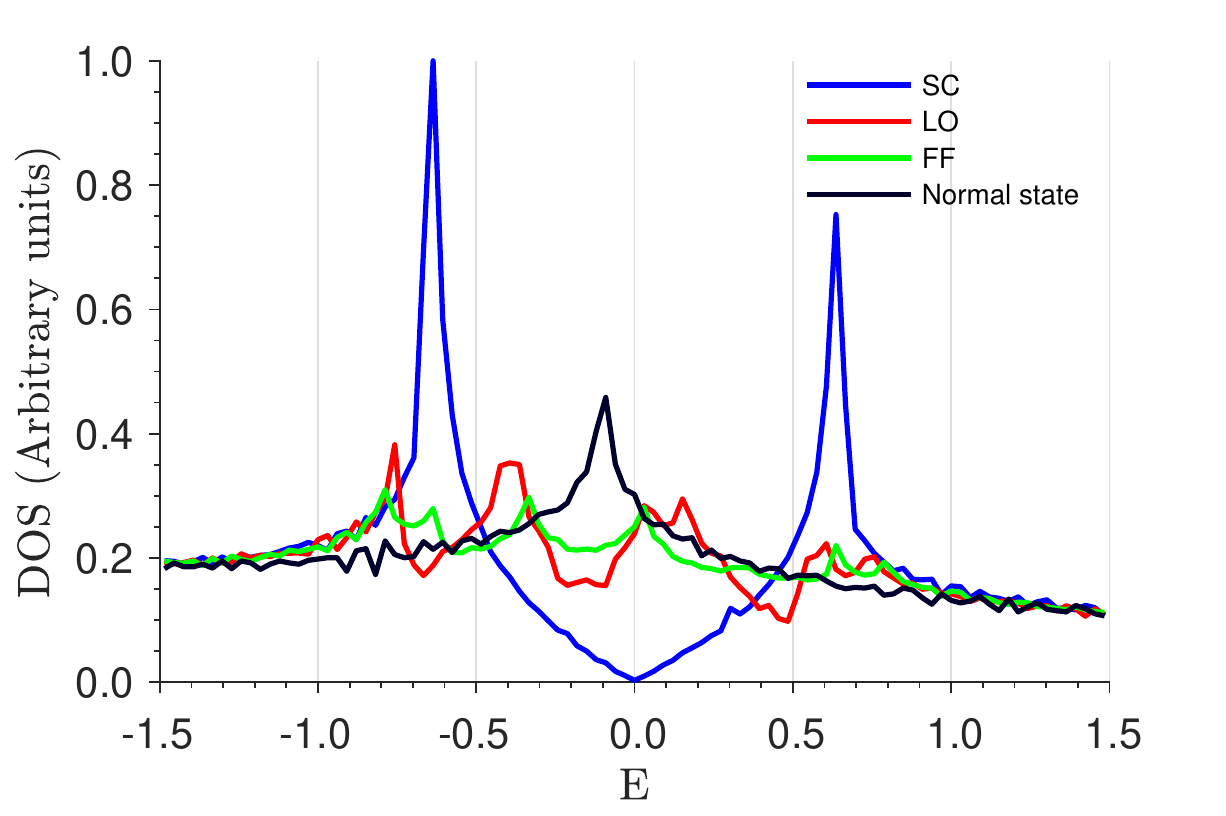}
	\caption{\label{DOS} (Color online) The density of states for LO and FF at $T=0,\rho=0.8$, note that LO is the true ground state at this doping. An ordinary d-wave SC is shown in comparison to the same parameters but with $v(\ve{q})=1$, as well as the corresponding normal state. }
\end{figure}

As comparison, the density of states (DOS) is included in FIG \ref{DOS} alongside the normal state and an ordinary d-wave SC at same coupling. The spectral function shows similar features as previous works \cite{baruch2008spectral,lee2014amperean,agterberg2015emergent} with a finite DOS at the Fermi-surface. Since the calculation of the spectral function is not dependent on the self-consistency of order-parameters we have not addressed these in detail here.

\section{\label{current_bloch}Current operator and Bloch's theorem}
In this section we will investigate cancellation of current in the FF state, predicted by Bloch's theorem on ground state current. Consider a uniform current $\ve{J}_{\ve{q}=0}$ derived from the continuity equation ${\lim}\atop{\ve{q}\rightarrow 0}$ $\ve{q}\cdot\ve{J}_{\ve{q}}=[H,\rho_\ve{q}]$ giving 
\begin{eqnarray}
\ve{J}&=&\sum_{\ve{k}, \sigma} v_{\ve{k}}n_{\ve{k}, \sigma}+\frac{1}{N}\sum_{\ve{k}, \ve{k}' ,\ve{q}} 2\left (\ve{\nabla}_{\ve{q}} V(\ve{k},\ve{k}',\ve{q})\right)\\
&& \times \, c^{\dagger}_{\uparrow, \ve{k}+\frac{\ve{q}}{2}}c^{\dagger}_{\downarrow, -\ve{k}+\frac{\ve{q}}{2}}c_{\downarrow, -\ve{k}'+\frac{\ve{q}}{2}}c_{\uparrow, \ve{k}'+\frac{\ve{q}}{2}} \nonumber \,.
\label{current}
\end{eqnarray}
(Here and subsequently, we drop the $\ve{q}=0$ subscript on $\ve{J}$ ). Alternatively, and with the same result, the expression for the current operator can be derived from a Peierls substitution of a constant vector potential in the lattice Hamiltonian, through $\ve{J}=\frac{d H}{d\ve{A}}|_{\ve{A}=0}$. The first term is the ordinary single-particle current operator, $\ve{J}_{\text{sing}}$, and the second is related to the pair-hopping and will be referred to as the anomalous current operator $\ve{J}_{\text{an}}$. The anomalous term vanishes in the limit where the interaction is $\ve{q}$-independent ($V(\ve{k},\ve{k}',\ve{q})=V(\ve{k},\ve{k}')$), which is the case for a density-density interaction. Bloch's theorem implies that the ground state of an interacting Hamiltonian cannot carry a current. This is readily shown using the polarization operator $\ve{P}=\frac{1}{N}\sum_i\ve{r}_in_i$, which (up to a total derivate) satisfies $\dot{\ve{P}}=\ve{J}$. Assuming a current carrying ground state $\ket{\phi}$, with current $\ve{J}_{\phi}$, we can construct another state $\ket{\psi}=e^{i \pmb{\delta} \cdot \ve{P}} \ket{\phi}$, with $\pmb{\delta}$ as a vector parameter. Evaluating the energy (assuming $T=0$) of this state to linear order in $\pmb{\delta}$ we find 
\begin{equation}
E_\psi=\bra{\psi} H \ket{\psi}=E_{\phi} +i \pmb{\delta} \cdot \bra{\phi} [H,\ve{P}] \ket{\phi} =E_{\phi}+\pmb{\delta} \cdot \ve{J}_{\phi}\,,
\label{bloch_proof}
\end{equation}
where $E_\phi=\bra{\phi} H \ket{\phi}$ and $\ve{J}_{\phi}=\bra{\phi} \ve{J} \ket{\phi}$. With a suitable choice of $\pmb{\delta}$ we can always lower the energy compared to a putative ground state with a current. The proof of this theorem is typically presented for a Hamiltonian with density-density interactions,\cite{bohm1949note,ohashi1996bloch} which trivially commutes with the polarization operator, but here we see that it holds more generally.
%\com{But, the interaction term \eqref{interaction} is not on a density-density term, and we see that the theorem hold more generally}. 
Also, note that the theorem applies to all locally stable states and not just the ground state.

A mean-field calculation is a minimization within a subspace of states and therefore it is not, a priori, obvious that it will respect Bloch's theorem. This problem was already discussed by \citet{fulde1964superconductivity}. However, by using the same arguments as before, we can show that Bloch's theorem will be obeyed also within the mean-field subspace. Focusing on the FF state with arbitrary momentum $\ve{Q}$ we have the explicit representation of the  BCS-like ground state at self-consistency
\begin{equation}
\ket{\ve{Q}}=\prod'_{\ve{k}\sigma}c^\dagger_{\ve{k}\sigma}\prod''_{\ve{k}}(u_k+v_kc^\dagger_{\ve{k},\uparrow}c^\dagger_{-\ve{k}+\ve{Q},\downarrow})\ket{0}\,,
\label{FF_ground}
\end{equation} 
where $u^2_k,v^2_k=\frac{1}{2}(1\pm\frac{\xi_{\ve{k}}+\xi_{-\ve{k}+\ve{Q}}}{\sqrt{(\xi_{\ve{k}}+\xi_{-\ve{k}+\ve{Q}})^2+4\Delta^2_{\ve{Q}}(\ve{k}-\ve{Q}/2)}})$. Here 
$\prod'$ is over momenta $\ve{k}$ such that $\xi_\ve{k}<0$, $\xi_{-\ve{k}+\ve{Q}}>0$ and $|\xi_{\ve{k}}\xi_{-\ve{k}+\ve{Q}}|>\Delta^2_{\ve{Q}}(\ve{k}-\ve{Q}/2)$ for which the quasi-particle has lower energy than the finite momentum Cooper pair, consequently $\prod''$ is over the remaining momenta. Assuming that this state has current $\ve{J}_\ve{Q}$ and energy $E_\ve{Q}$ we construct a new state  
\begin{eqnarray}
&&\widetilde{\ket{\ve{Q}+2\delta\ve{Q}}}=e^{i \delta\ve{Q} \cdot \ve{P}} \ket{\ve{Q}}=\prod'_{\ve{k} \sigma}c^\dagger_{\ve{k}+\delta\ve{Q}\sigma}\nonumber\\
&\times&\prod''_{\ve{k}}(u_k+v_kc^\dagger_{\ve{k}+\delta\ve{Q},\uparrow}c^\dagger_{-\ve{k}+\bold{Q}+\delta\ve{Q},\downarrow})\ket{0}\,,
\end{eqnarray} 
where $v_k$ and $u_k$ are unaffected by the shift. To linear order it is easy to show that this state has energy $E_{\ve{Q}+\delta\ve{Q}}=E_\ve{Q}+\delta \ve{Q}\cdot \ve{J}_\ve{Q}$, indeed in accordance with \eqref{bloch_proof}. Thus the energy can be lowered by an appropriate small shift of the pair-momenta anti-parallel to the current. The state $\widetilde{\ket{\ve{Q}+2\delta\ve{Q}}}$ is not a self-consistent solution at pair-momentum $\ve{Q}+2\delta\ve{Q}$, but it has the correct particle number (inherited from the unshifted state, given by $n=\sum'_k2+\sum''_k2v^2_k$) and is contained in the variational space of the Hamiltonian at $\ve{Q}+2\delta \ve{\ve{Q}}$, so its energy will be higher than the corresponding self-consistent state. Thus by successive iterations of momentum-shift and convergence to self-consistency, we can continue to lower the energy until the ground state is found. Hence the mean-field calculation is true to Bloch's theorem. 
\begin{figure}[h!]
	\centering
		\includegraphics[width=0.5\textwidth]{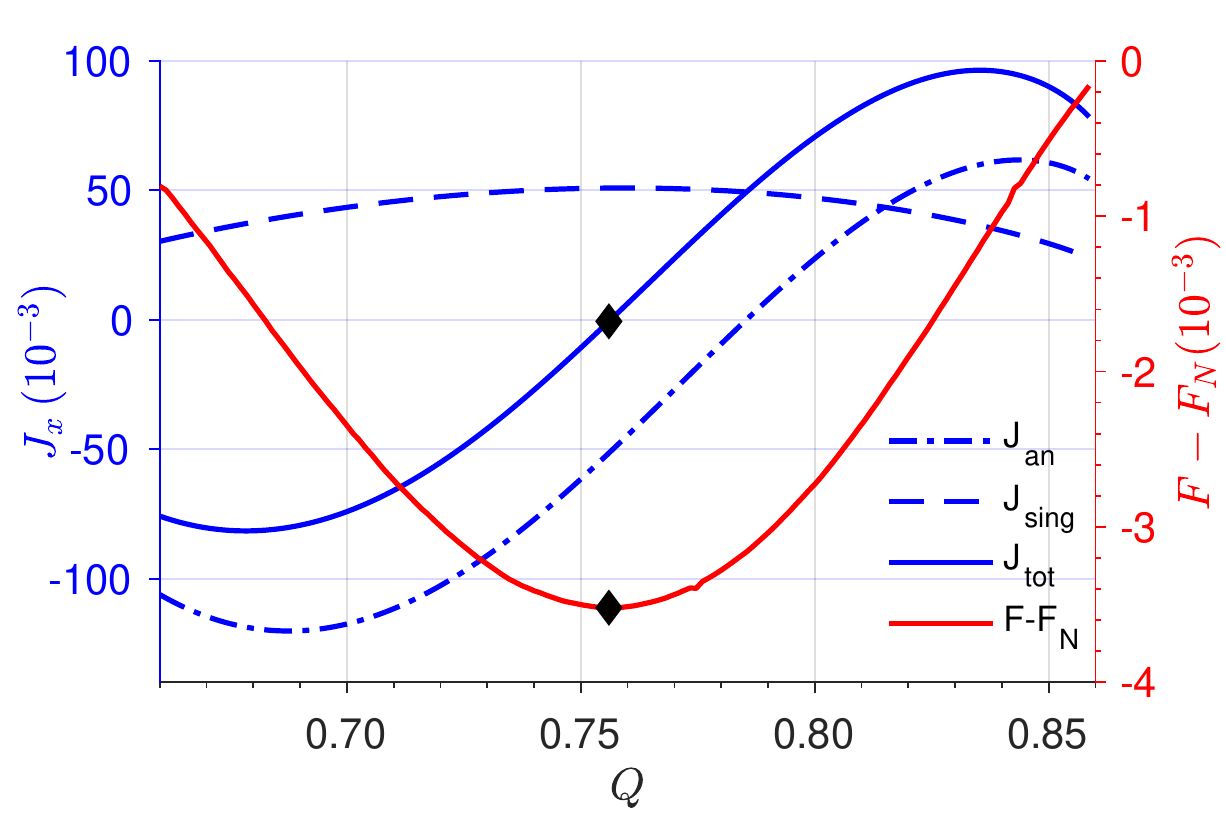}		
	\caption{\label{FF_bloch} (Color online) Free energy(red) and current(blue) for an FF state ($\Delta_{\ve{Q}_1} \neq 0 $,$\Delta_{\ve{Q}_2} = 0 $) as a function of $\ve{Q}_1=Q \hat{x}$ at $T=0.02, \rho =0.65$. The single-particle current (dashed) is always positive for $Q>0$ while the anomalous current (dot-dashed) cancels the former exactly at the minimum in energy, marked by diamonds. }
\end{figure}

In FIG \ref{FF_bloch} we show the energy and current 
as a function of $\ve{Q}_1=Q \hat{x}$ for the FF ground state with $\Delta_{\ve{Q}_1} \neq 0 $. The anomalous current, $\ve{J}_\text{an}$, exactly compensates the single-particle current, $\ve{J}_\text{sing}$, at the energy minimum in accordance with Bloch's theorem. Stated differently: Since $\ve{J}_\text{sing}$ is everywhere positive, the anomalous current $\ve{J}_\text{an}$ is a necessity to make it a valid ground state. 

As a consequence, any time-reversal breaking ground state must have a mechanism for cancellation of current. We have seen that pair-hopping yields such a mechanism through the introduction of an anomalous current. Another mechanism, which does not require an anomalous term, is to consider the original Fulde-Ferrell state which arises because of spin-population imbalance.\cite{fulde1964superconductivity} In this case, the single-particle term cancels itself because of the counter-propagating quasi-particle excitation current. However, a system with spin-population imbalance explicitly breaks time-reversal symmetry, while our system breaks it spontaneously. In the ordinary nearest-neighbor attraction model without anomalous current or population imbalance, it seems not possible for FF to be the ground state (see Appendix \ref{nearest_finite_q}).

 \section{Current in a superconductor \label{current_section}}
  
We will now turn to the LO state and discuss how it is affected by a homogeneous current (for details about the physical relevance of this approach see Appendix \ref{appendix_homogenous_current}). We will first review the procedure in an ordinary superconducting state with one order-parameter, and then discuss the necessary generalizations in order to consider current in a PDW-state with two order-parameters.

The usual procedure\cite{khavkine2004supercurrent,kee2004critical} is to construct a finite momentum condensate $\Delta_{\ve{q}_s} = \av{c_{\downarrow, \ve{k} + \frac{\ve{q}_s}{2}}c_{\uparrow, -\ve{k} + \frac{\ve{q}_s}{2}}}$, which carries a current $\ve{J}_c  = \frac{e n}{m} \ve{q}_s$. As a result the spectrum becomes Doppler-shifted\cite{bardeen1962critical,maki1962quantization,maki1963persistent,tinkham1996introduction} $ E=\nabla_k \varepsilon \cdot \ve{q}_s+\sqrt{(\varepsilon - \mu)^2+|\Delta|^2}$ and states with $E<0$ will be excited already at $T=0$. These excited states constitutes a counter propagating current $\ve{J}_e$, and we write the total current as $\ve{J} = \ve{J}_c - \ve{J}_e$. The destruction of the total current as a function of $\ve{q}_s$ is two-fold: (i) More quasi-particles will be excited, which enhances $\ve{J}_e$, and (ii) excitations will change the self-consistent condition and deplete the condensate, resulting in a smaller $\ve{J}_c$. In general the function $\ve{J}(\ve{q}_s)$ will be concave\cite{khavkine2004supercurrent,kee2004critical} %(with possible discrete jumps)
and one identifies the \textit{depairing} current, $\ve{J}_d$, with the greatest possible current $ \nabla_{\ve{q}_s} \ve{J} \mid_{\ve{J}_d}=0$. 

This procedure can be understood within the mean-field construction outlined in Section \ref{mean-field}, where one considers different mean-field Hamiltonians parametrized by $\ve{q}_s$, which is varied to find the lowest energy state with a specific current. For the ordinary superconductor, one obtains a mapping, $\ve{q}_s \rightarrow \ve{J}$, consisting of two branches (i.e. two $\ve{q}_s$ correspond to the same $\ve{J}$) and one picks the lowest energy branch. 

For PDW we proceed in a similar manner by considering the effective Hamiltonian $H_{\text{MF}}$ in \eqref{eff_H}, but with $\ve{Q}_1 = \ve{Q}_0 + \ve{q}_{1}, \ve{Q}_2 = -\ve{Q}_0+ \ve{q}_{2}$, shifted from the minimum. Thus we are considering the mapping $\ve{q}_1,\ve{q}_2 \rightarrow \ve{J}$, which is no longer just doubly degenerate but the degeneration is spanned by two continuous dimensions. Nevertheless the procedure is in principle straightforward: We should minimize the energy over a subset of $\ve{q}_1,\ve{q}_2$ which corresponds to one specific current $\ve{J}$. 

In the remainder of this paper, we will focus on systems with LO ground state picking $\rho = 0.8$. We need to probe all states spanned by $\ve{Q}_1, \ve{Q}_2$. Even though the ground state is an LO state the current carrying metastable state might not be, thus we will consider FF and LO states independently. (Note that we extend the meaning of LO to include $\Delta_{\ve{Q}_1} \neq \Delta_{\ve{Q}_2}, |\Delta_{\ve{Q}_{(1,2)}}| >0  $.)
Because of the rather complex minimization problem, we do not aim to fully investigate the various parameter-dependent possibilities, instead we focus on two features characteristic of the PDW state:  A \textit{phase-separation} between LO and FF solutions and spontaneous \textit{mirror-symmetry breaking}. 

\subsection{\label{phase_sep} Current induced phase-separation }

From a symmetry perspective, driving current transverse to $\Q= \sQ \hat{x}$ is similar to an ordinary superconductor since the state is expected to be invariant under order-parameter exchange, $\Delta_{\ve{Q}_1} \leftrightarrow\Delta_{\ve{Q}_2}$. In contrast, a state with the current along $\Q$ is not symmetric under order-parameter interchange and we expect this case to be fundamentally different, hence, current along $\Q$ will be the main target of the investigation.
\begin{figure}[h!]
	\centering
	\includegraphics[width=0.5\textwidth]{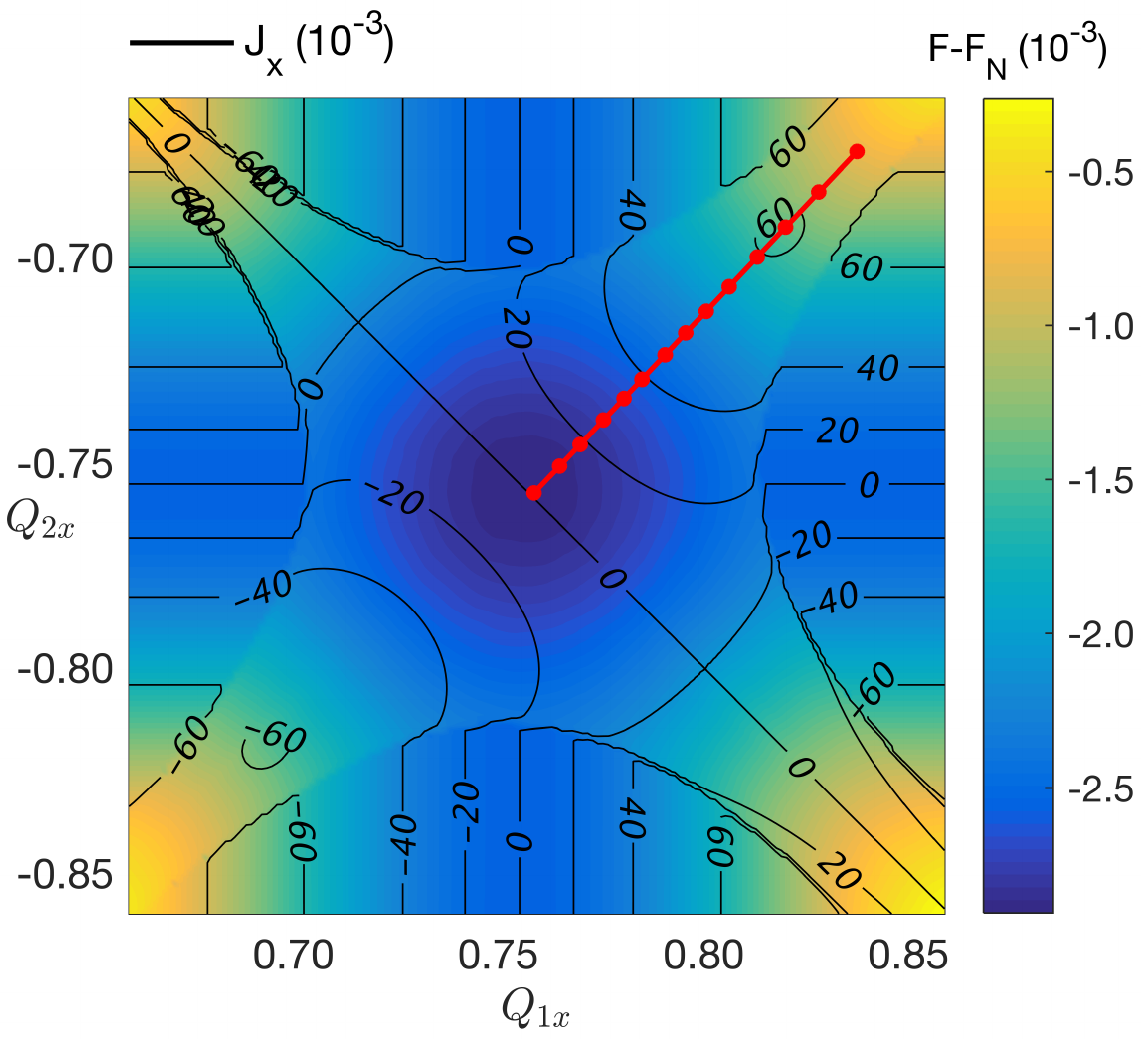} %
	\caption{\label{J_x_stable}  (Color online) The free energy $F$ of the LO state (compared to the normal state energy $F_N$) as a function of $Q_{1x}$ and $Q_{2x}$ for $T=0.02$, $\rho = 0.8$. The solid black lines are solutions of equal current. The red dots correspond to the extreme energy values for each current, also shown in FIG \ref{J_x_energy}. Note that the LO state is only stable in the center region where the energy forms circular equipotent contours, in the outer regions it decays to an FF state, indicated by the independence of either $Q_{1x}$  or  $Q_{2x}$. }
\end{figure}

 As mentioned, the problem with considering two order-parameters is the extended parameter space. In general, we can construct a state for each pair of $\ve{Q}_1,\ve{Q}_2$, however, we are only interested in the lowest energy state for each current, which reduces the relevant parameter space to a two-dimensional subspace of $\ve{Q}_1,\ve{Q}_2$. By considering current along $\Q$ we may confine to the parameterization $\ve{Q}_1 =(Q_{1x},0), \ve{Q}_2 = (Q_{2x},0)$, which only induces current along $x$. There are however additional states that only have current along $x$ which we can construct by shifting $\ve{Q}_1 $ and $\ve{Q}_2$  in opposite $y$-directions. These states would break mirror-symmetry $\ve{Q}_{(1,2)y} \rightarrow -\ve{Q}_{(1,2)y}$ and we explore this possibility in Section \ref{current_all_angles}. Here we prevent this symmetry breaking by picking $\kappa_y$ sufficiently small, making this shift energetically expensive.  (This assumption was confirmed for $\kappa_y = \kappa_x=0.2$.)

\begin{figure}[h!]
	\centering		
	\includegraphics[width=0.5\textwidth]{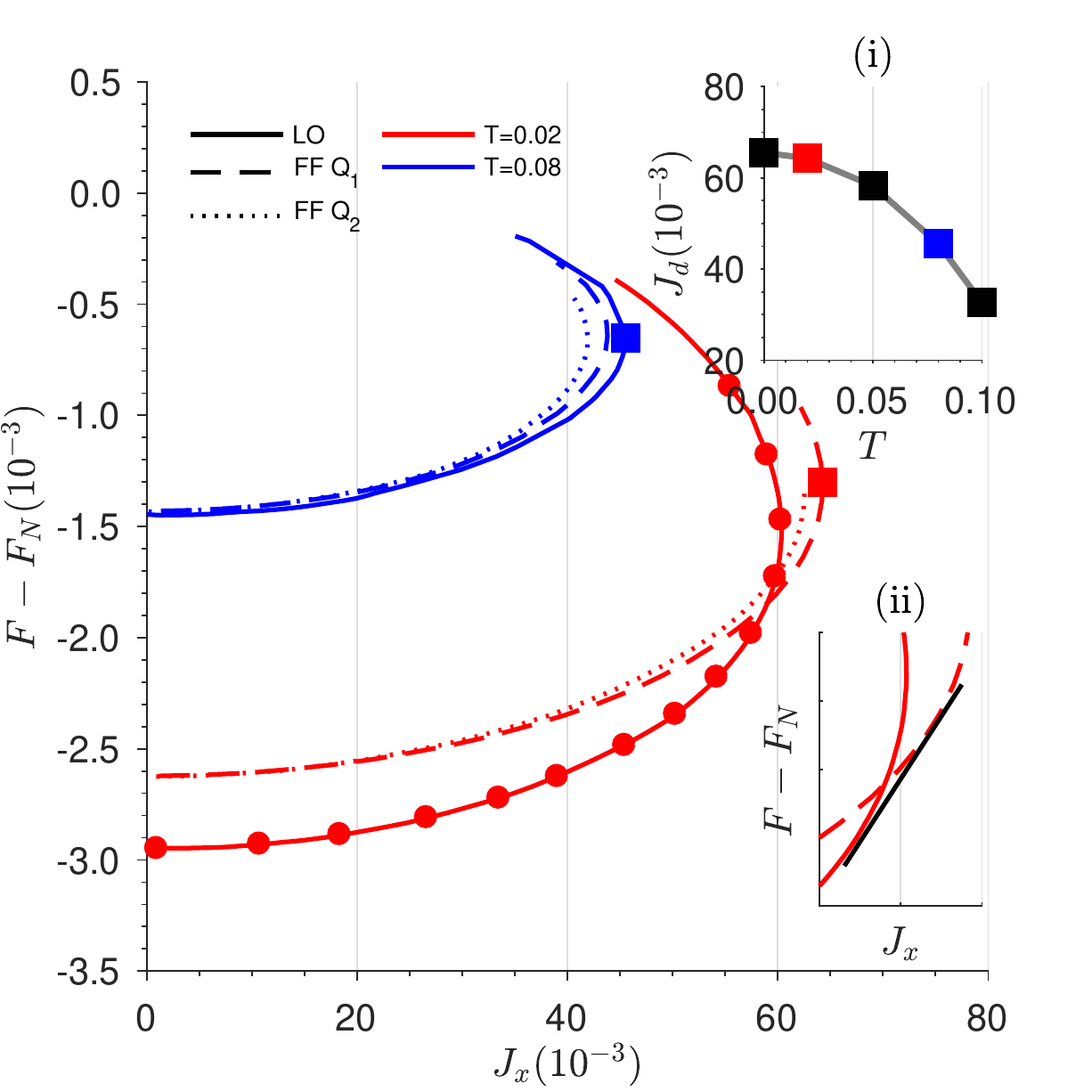}
	\caption{\label{J_x_energy} (Color online) Lowest energy solution as a function of current along $\Q=\sQ \hat{x}$ for $T=0.02\, (\text{red}) $ and $0.08 \, (\text{blue})$, $\rho = 0.8$ (and for small enough $\kappa_y$, see text). Solid (dashed) lines represents LO (FF) states. FF $\ve{Q}_1$ ($\ve{Q}_2$) refers to $\ve{Q}_1 \neq 0$ ($\ve{Q}_2 \neq 0$). The circles represents the same solution as in FIG \ref{J_x_stable}  and squares the depairing current. Inset (i): Temperature dependence of the depairing current. Inset (ii) : A magnified graph over the LO to FF transition; the lowest energy solution is a phase-separated one with weight in both LO and FF, indicated by the intersection of the black line.}
\end{figure}

In FIG \ref{J_x_stable} the free energy for the LO states as a function of the order-parameter momenta is presented. The energy forms circular equipotent contours for $Q_{1x}, Q_{2x}$ near the energy minimum (at $ \pm Q_0$), in this region LO is stable and both order-parameters are non-zero. When increasing the momenta further the LO state becomes unstable towards an FF state. These FF states are independent of $Q_{1x}\, (Q_{2x})$ indicating that $\Delta_{\ve{Q}_1}=0 \,(\Delta_{\ve{Q}_2}=0)$. (Note that there exist FF states of higher energy even where the LO state is stable.) The extreme energy states for a specific current in positive $x$-direction is traced out in red dots (also shown in FIG \ref{J_x_energy}).

The extreme energy LO and FF solutions as a function of current are presented in FIG \ref{J_x_energy} for $T=0.02; 0.08$ where we identified the lowest energy solution for each current. Note that we include two types of FF states with $\Delta_{\ve{Q}_1} \neq 0,\Delta_{\ve{Q}_2} = 0$ and $\Delta_{\ve{Q}_1}=0,\Delta_{\ve{Q}_2} \neq 0$ respectively. At higher temperatures LO dominates over FF for all currents, however, they also become more degenerate. For low temperatures the FF state survives for higher current and near depairing current (shown as a function of temperature in \ref{J_x_energy}(i)) the system makes a transition from LO to FF. In this region the energy is concave, highlighted in inset \ref{J_x_energy}(ii), which suggest a phase-separation, indicative of a first order transition. The lowest energy configuration for a state in this interval
%in the interval marked by the black line 
is given by a superposition of an LO and FF state situated at the intersection with the black line, in accordance with Maxwell's construction. The phase-separation is not very strong in this system, but since it arises from near degenerate FF and LO solutions we expect it to be a generic feature in systems with this property.
\begin{figure}[h!]
	\centering		
	\includegraphics[width=0.5\textwidth]{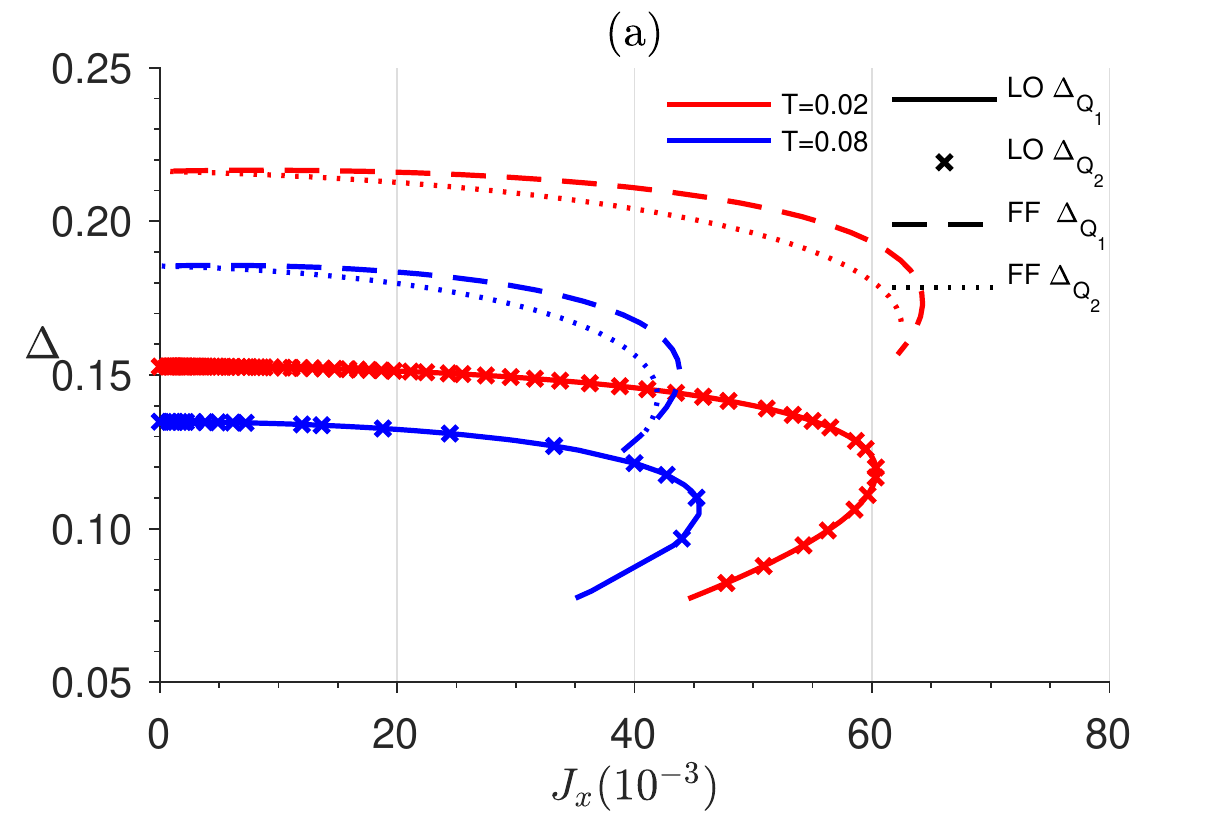}
	\\
	\includegraphics[width=0.5\textwidth]{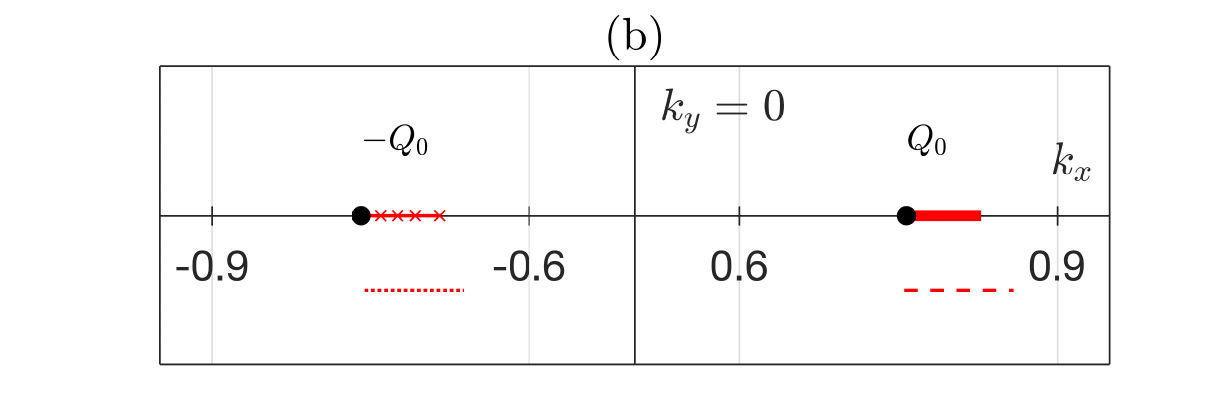} %
	\caption{\label{J_x_order} (Color online) (a):  The order-parameters for lowest energy solution as a function of current along $\Q=\sQ \hat{x}$ for $T=0.02 \, (\text{red}) $ and $0.08 \, (\text{blue})$, $\rho = 0.8$ (same solution as in FIG \ref{J_x_energy}). FF $\Delta_{\ve{Q}_1} (\Delta_{\ve{Q}_2})$ refers to different states $\Delta_{\ve{Q}_1} \neq 0,\Delta_{\ve{Q}_2} = 0$ $(\Delta_{\ve{Q}_1}=0,\Delta_{\ve{Q}_2} \neq 0$) while LO $\Delta_{\ve{Q}_1} (\Delta_{\ve{Q}_2})$ refers to the same state. (b): Evolution of $\ve{Q}_1$ (rightmost side) and $\ve{Q}_2$ (leftmost side), note that $Q_{(1,2)y}=0$ and the displacement between FF and LO in $y$-direction is just for distinguishability.
	}
\end{figure}

In FIG \ref{J_x_order}(a) the evolution of the order-parameters is presented as a function of current. The FF and LO state shows similar behavior, however, FF carries more weight in its non-vanishing component than LO does in each of its components. (Note that FF $\Delta_{\ve{Q}_1} (\Delta_{\ve{Q}_2})$ refers to different states $\Delta_{\ve{Q}_1} \neq 0,\Delta_{\ve{Q}_2} = 0$ $(\Delta_{\ve{Q}_1}=0,\Delta_{\ve{Q}_2} \neq 0$) while LO $\Delta_{\ve{Q}_1} (\Delta_{\ve{Q}_2})$ refers to the same state.) One might expect that the LO state would enhance the condensate running in the forward directing and deplete the trailing one in order to create a current running forward. Surprisingly we see that orders remain (to high accuracy) of same size, albeit $\ve{Q}_1 \neq -\ve{Q}_2$ as is seen in \ref{J_x_order}(b). It seems as the LO state can utilize the anomalous current from the pair-hopping, which is equally benefited by both orders. Without pair-hopping this feature is expected to change and the phase-separation is likely not to occur since the near degeneracy between FF and LO is expected to be lifted (see Appendix \ref{nearest_finite_q}). In the next section we will see a different scenario emerge, where instead two distinct LO states are near degenerate.

For reference, currents are measured in units of $\frac{e}{\hbar} \frac{t}{ac}$ with $t,a,c$ being the hopping parameter and lattice axes respectively. As an order-of-magnitude estimate we find $J_{d,x} \backsimeq 2.0 \times 10^8\,$A$/$cm$^2$ (with $t = 350 \, $meV and $a,c$ appropriate for YBCO\cite{benzi2004oxygen}) which is in line with measured values.\cite{nawaz2013microwave}

%Making superficial contact with experiments we note that the currents are measured in units of $\frac{e}{\hbar} \frac{t}{ac}$ with $t,a,c$ being the hopping parameter, and lattice axes respectively. With $t = 350 \, $meV and $a,c$ appropriate for YBCO\cite{benzi2004oxygen} we find $J_{d,x} \backsimeq 2.0 \times 10^8\,$A$/$cm$^2$ which is in line with measured values.\cite{nawaz2013microwave}

\subsection{\label{current_all_angles} Mirror-symmetry breaking}

\begin{figure}[h!]
	\centering
	\includegraphics[width=0.5\textwidth]{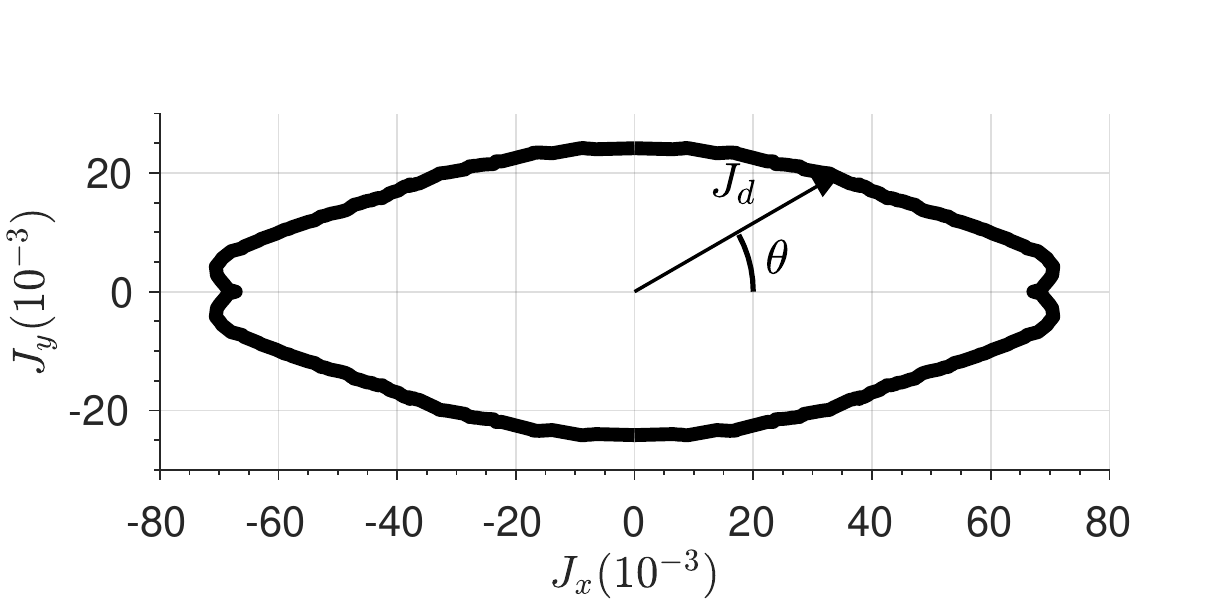}
	\caption{\label{all_angles} The directional dependence of the depairing current with $\kappa_{y} \rightarrow \infty$. The highest current solution for each angle was extracted. (Calculations were done for angles in the first quadrant, and the results was symmetrized.)}
\end{figure}

In this section we remove pair-hopping transverse to $\Q=\sQ \hat{x}$, corresponding to $\kappa_y \rightarrow \infty$. In FIG \ref{all_angles} the directional dependence of the depairing current is shown and we note a rather high anisotropy of $J_{d,x}/J_{d,y} \simeq 3.2$. 

\begin{figure}[h!]
	\centering
	\includegraphics[width=0.23\textwidth]{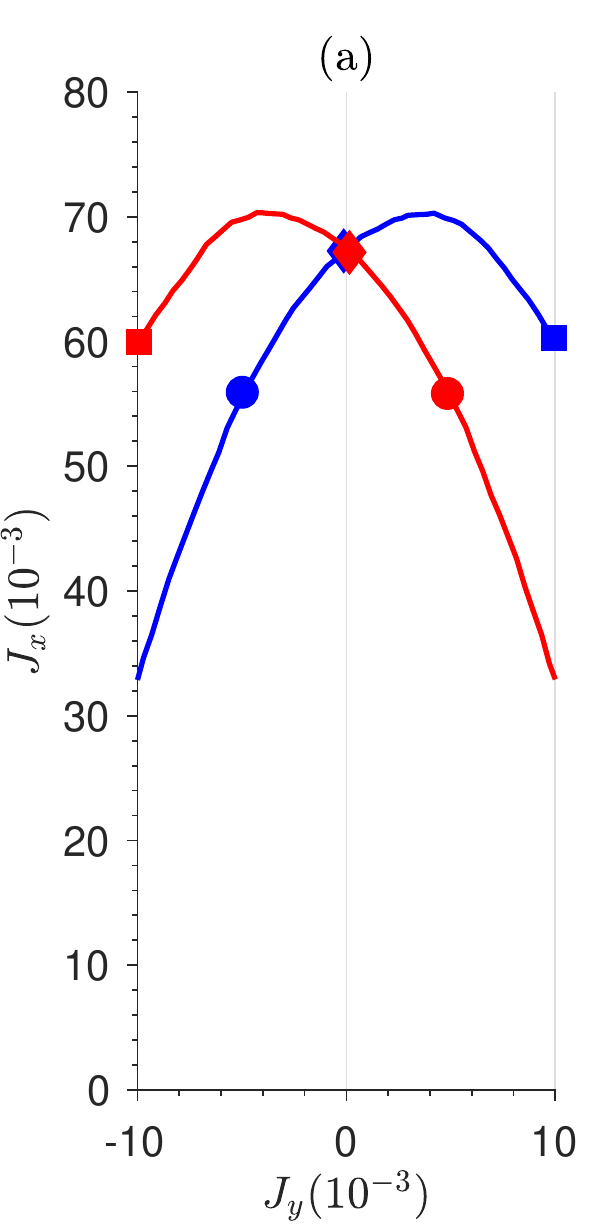}
   \includegraphics[width=0.23\textwidth]{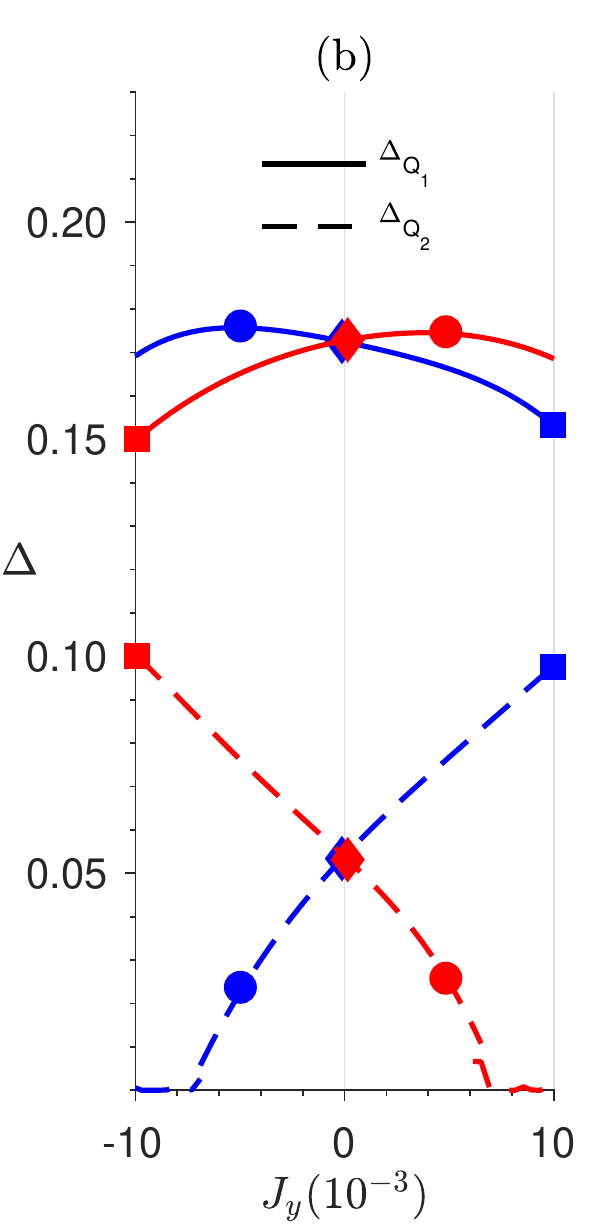}		
   \\
 \includegraphics[width=0.5\textwidth]{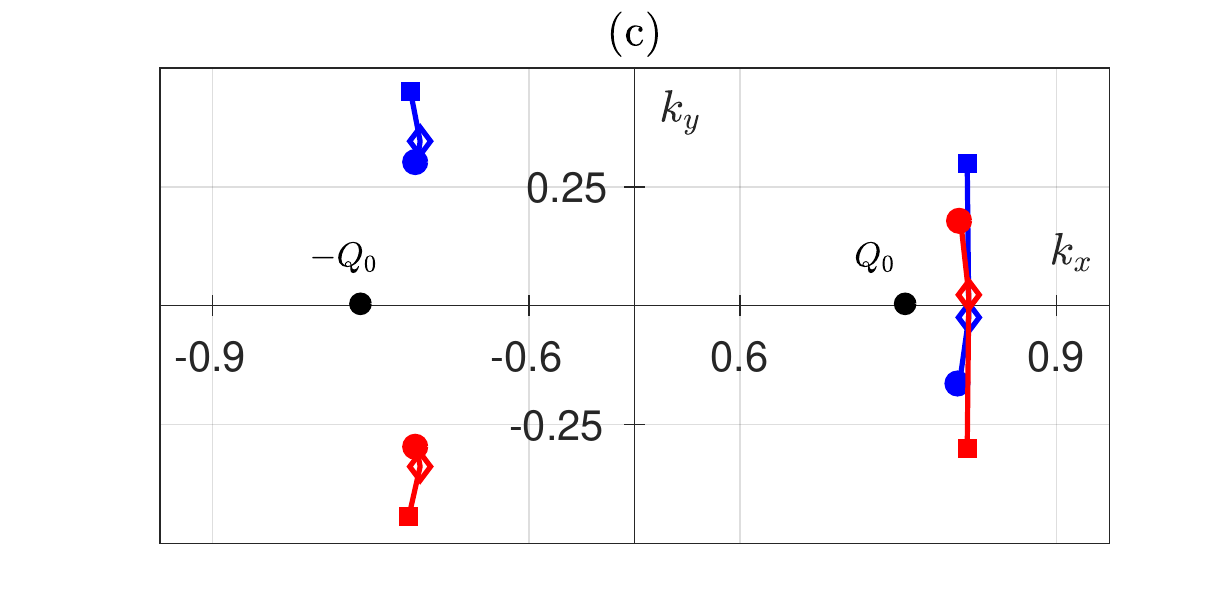}	  
	\caption{\label{cusp} (Color online) Highest current solution along $\Q=\sQ \hat{x}$ separated into the mirror-symmetry connected branches $Q_{2y}>0$ (blue) and $Q_{2y}<0$ (red) ($\kappa_{y} \rightarrow \infty$). Square and circle markers indicate the end of the interval shown in (c) while the diamond markers indicate $J_y=0$. In (a) we see the emergent cusp at $J_y=0$ where both mirror-symmetric branches meet. (b): The evolution of the corresponding order-parameters. (c): The evolution of $\ve{Q}_1$ (rightmost side) and $\ve{Q}_2$ (leftmost side). (The data was interpolated over a denser $\ve{Q}_{(1,2)}$ grid and (b) was fitted to a 4th degree polynomial with maximum error $0.007$ in $\Delta_{\ve{Q}_{(1,2)}}$.)}
\end{figure}

More striking is the apparent cusp at $\theta = 0$, which is highlighted in FIG \ref{cusp}(a,b). To understand where this comes from we look at two subbranches of the solution, one where $Q_{2y}>0$ and one where $Q_{2y}<0$. The highest $J$ for every angle is shown for each branch in FIG \ref{cusp}(a). In contrast to what we saw in Section \ref{phase_sep} we can optimize current along $\Q$, (i.e. in $x$-direction) by picking finite values of $Q_{(1,2)y}$. In FIG \ref{cusp}(c) the evolution of $\ve{Q}_{1}$ (rightmost curves) and $\ve{Q}_{2}$ (leftmost curves) are shown, we see that the two branches are mirror-symmetric partners that map to each other under $Q_{(1,2)y} \rightarrow -Q_{(1,2)y}$. The cusp results from the crossing of the two branches, see FIG \ref{cusp}(a). In FIG \ref{cusp}(b) we see that $|\Delta_{\ve{Q}_1}| >|\Delta_{\ve{Q}_2} |$, thus the condensate running in the forward direction is promoted, in contrast to FIG \ref{J_x_order}(a). 

At $J_y=0$ there are two degenerate solutions, one from each branch. $J_y$ cancel since the difference in the order-parameters $|\Delta_{\ve{Q}_1}|>|\Delta_{\ve{Q}_2}|$ (see FIG \ref{cusp}(b)) are compensated by a corresponding difference in momenta $|\ve{Q}_{1y}| < |\ve{Q}_{2y}|$. In fact, increasing $J_x$ slightly, the system spontaneously picks one of the branches and starts to conduct current along $y$ as well. This induced current along $y$ is conceptually similar to what was found by \citet{doh2006novel} (for a quite different system) who also proposed an experimental test of this property by putting the system on a cylindrical geometry (with $y$ in the circular direction) and measuring flux induced by $J_y$.

\section{\label{discussion}Summary and Outlook }
We have studied an extension of the standard effective model of d-wave superconductivity in the cuprate superconductors where the nearest-neighbor attraction is extended into a modulated longer range pair-hopping. The pair-hopping is such that it only breaks the rotational symmetry of the lattice, and is modulated by a wave vector $\Q=\sQ\hat{x}$ (aligned with a crystal axis).  By construction, it suppresses uniform superconductivity while promoting the formation of a pair-density wave (PDW) with a periodically modulated superconducting order. We have used the model to make a self-consistent exploration of the space of two finite momentum order-parameters $\Delta_{\ve{Q}_1}$ and $\Delta_{\ve{Q}_2}$. Interestingly, there is a close competition between two different types of ground states: a Larkin-Ovchinnikov (LO) state with $\Delta_{\ve{Q}_1}=\Delta_{\ve{Q}_2}$ and $\ve{Q}_1=-\ve{Q}_2\approx\Q$ (the sought for PDW), and a Fulde-Ferrell (FF) state with only one finite $\Delta_{\ve{Q}}$ ($\ve{Q}\approx \pm\Q$). The former spontaneously breaks translational symmetry while the latter spontaneously breaks time-reversal and parity. Which state has lower energy depends sensitively on the detailed form of the interaction and parameters of the model. A quantum critical point separating the two solutions as a function of doping may be realized.

An important aspect of the analysis is an ``anomalous'' term in the static current operator that is due to the pair-hopping, and instrumental to a proper consideration of Bloch's theorem on no ground state current. It is this term that enables a BCS ground state given by a single component finite momentum condensate, i.e. the FF state, without spin-population imbalance. This in contrast to a density-density type attraction for which we find, as expected, that FF states are only metastable with a finite current. We expect that this observation should be pertinent to the suggestions of an FF-like state related to loop current order by Agterberg {\em et al}.\cite{agterberg2015emergent}

By finding the self-consistent $\Delta_{\ve{Q}_1}$,$\Delta_{\ve{Q}_2}$ for any $\ve{Q}_1$,$\ve{Q}_2$ one can explore the effect of uniform current in this model and a number of interesting scenarios emerge. In a parameter range where the LO state is the ground state, it may happen that the largest current can be carried by an FF state, % with the latter giving the maximum, 
giving the depairing current. For a range of currents close to the depairing current it turns out in this case that there is phase-separation; a smaller current LO type state coexisting with a larger current FF state. In a real system, and ideally, this could presumably manifest itself as a filamentary phase-separation along the direction of the current. 

Another interesting scenario, occurring in a different parameter regime, is that there are two degenerate LO type states, related by mirror-symmetry with respect to the crystal axis $x$, that carry the highest currents. Here there is a cusp in the depairing current as a function of angle with respect to this axis, as the system switches from one to the other of the mirror-symmetry partners.   
The depairing current along $x$ is actually larger when a small transverse current component is allowed. Thus in a cylindrical geometry, arranged such that the transverse current is free to flow around the cylinder, there would be a spontaneous circumferential current when the current along the axis is sufficiently large.

Translational symmetry breaking manifested by CDW order seems quite generic in cuprate superconductors. This observation would rule out the possibility of a pure FF state that was studied in this paper. Nevertheless, it is still conceivable that such a pairing state is FF-like in the sense of breaking time-reversal and parity, but with a secondary pair-amplitude generated by the coupling to the CDW. Another aspect of the model that should also be explored are the effects of nematic distortion of the single-particle energies, which may change aspects of the interplay between LO and FF ground states.

\section{Acknowledgements}
Calculations were performed on resources at Chalmers Centre for Computational Science and Engineering (C3SE) provided by the Swedish National Infrastructure for Computing (SNIC). We thank F. Lombardi, T. Bauch, J. Tranquada, and S. Kivelson for discussions.  

\appendix

\section{\label{nearest_finite_q} LO state in ordinary nearest-neighbor interaction}
 \citet{loder2010} reported that one can obtain a finite momentum condensate from an ordinary nearest-neighbor interaction (setting $T(\ve{r}_{ij}^{\LCp} \LCm \ve{r}_{kl}^{\LCp}) = \delta_{\ve{r}_{ij}^{\LCp},\ve{r}_{kl}^{\LCp}}$ in \eqref{interaction}) with interaction strength about $V=2.2$, however we find the actual value to be $V \gtrsim 6 $.     The calculation is at fixed particle number, thus the chemical potential varies over the range of paring momenta. This was wrongly accounted for in Fig. 1  of \citet{loder2010}, rather the value $-\mu N$ needs to be subtracted from each point of the curve (as discussed in Appendix \ref{appendix_mean_field}). The authors of
	this paper have confirmed that this shift was indeed not included. \footnote{Private communication.}

\begin{figure}[h!]
	\centering
	\includegraphics[width=0.5\textwidth]{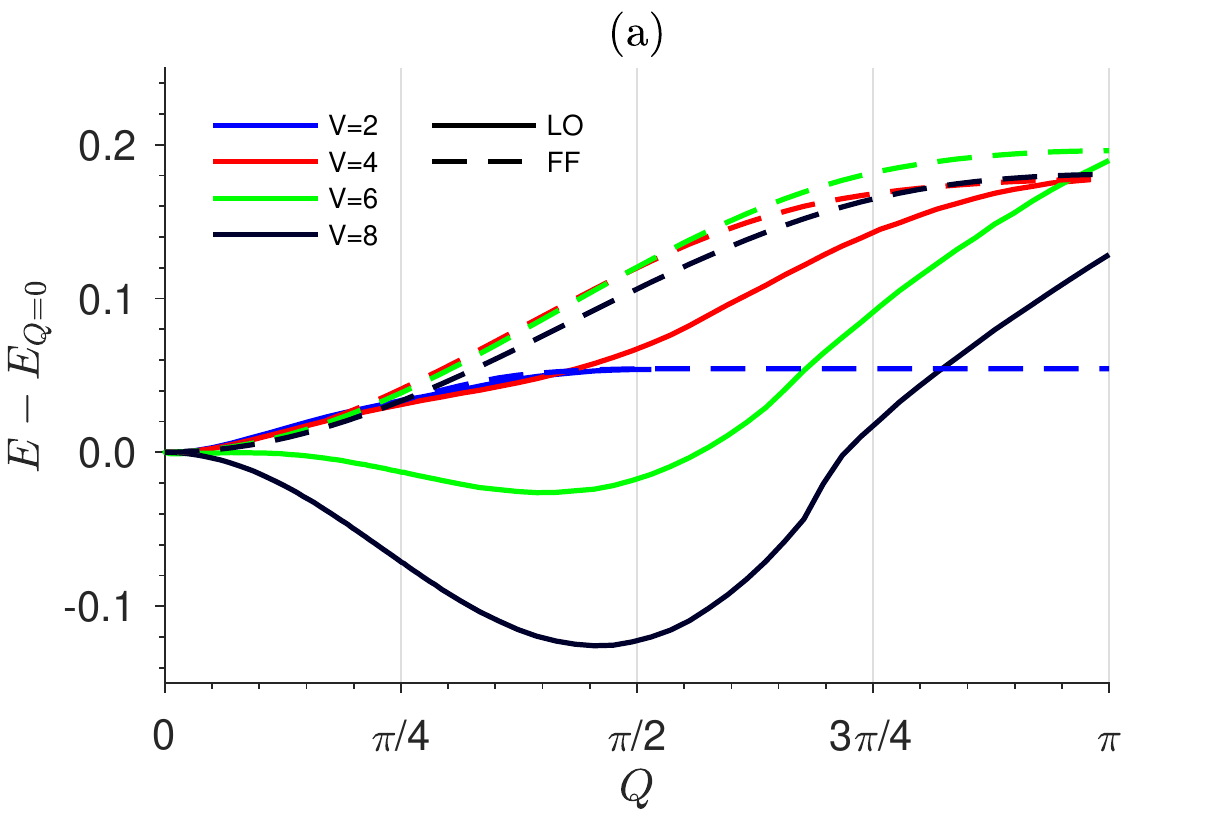} 
	\\
	\includegraphics[width=0.23\textwidth]{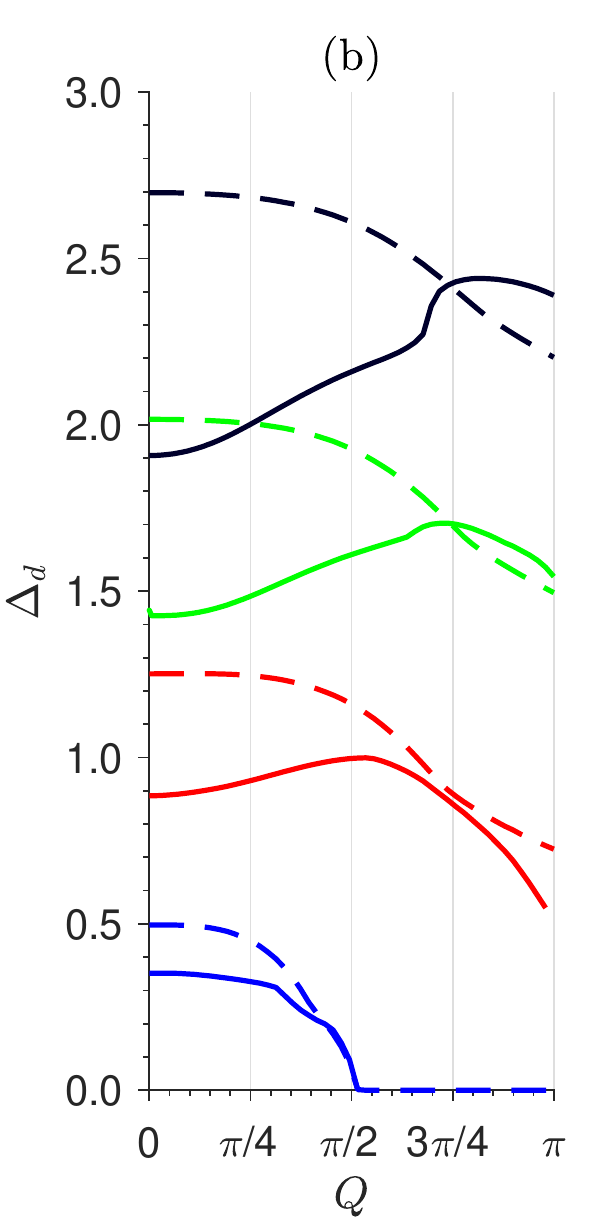} 
	\includegraphics[width=0.23\textwidth]{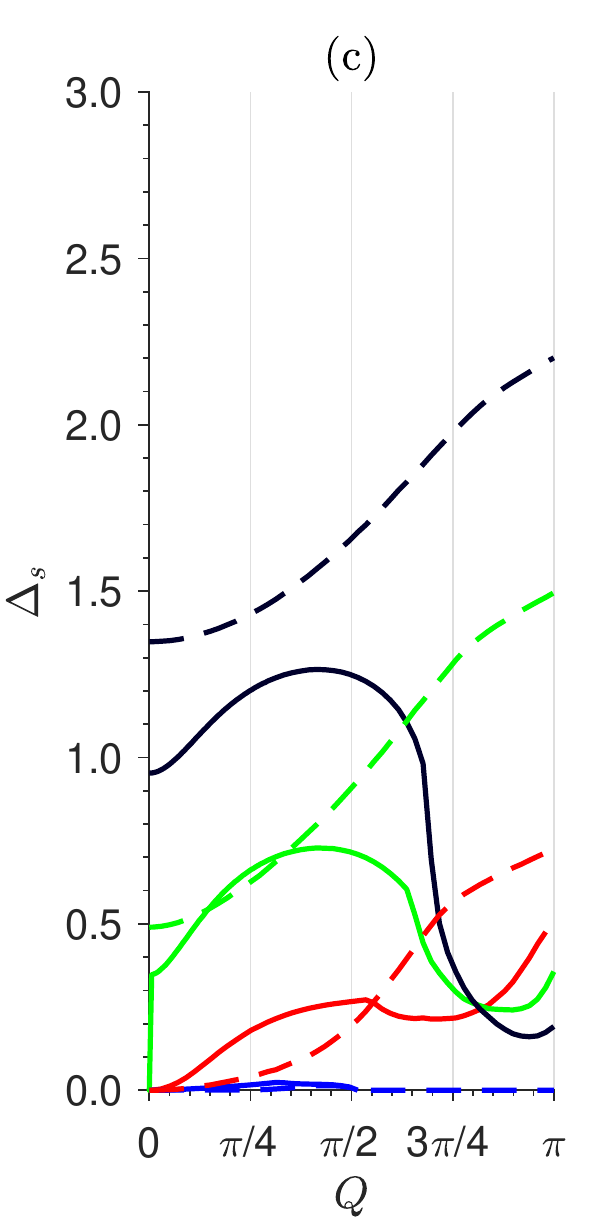} 
	\caption{\label{nearest} (Color online) Nearest-neighbor interaction model  ($T(\ve{r}_{ij}^{\LCp} \LCm \ve{r}_{kl}^{\LCp}) = \delta_{\ve{r}_{ij}^{\LCp},\ve{r}_{kl}^{\LCp}}$ in \eqref{interaction}) with finite momentum $\ve{Q}=Q \hat{x}$ for FF and LO states at $T=0$. (a): Energy as a function of $Q$, LO with finite momenta becomes the ground state for $V \gtrsim 6 $. (b),(c): D and s-wave order-parameter respectively, for higher interactions the s-wave component becomes substantial. }
\end{figure}

In FIG \ref{nearest} we show the corrected graphs using the same method as described above. In \ref{nearest}(a) we see the momentum dependence on the total energy and we note that an LO ground state indeed occurs but for interactions $V \gtrsim 6 $, which is far outside the weak coupling regime. In \ref{nearest} (b),(c) we present the d and s-wave order for corresponding momentum and interaction. First, we see that the order-parameters are very big and will heavily deform the Fermi-surface, secondly, we also acquire a substantial extended s-wave order. 

A final remark is that the FF state never becomes the ground state, instead it monotonically increases with $Q$. 
Thus, if we consider a Landau-Ginzburg type formulation of the free energy in terms of two order-parameters with opposite momenta, we see that it is the attraction between the two that stabilizes the LO state. This is in contrast to the pair-hopping model, that has been the main focus of this paper, where each finite momentum component may order independently. The interaction between the two may be attractive or weakly repulsive, corresponding to an LO ground state, or highly repulsive, leading to an FF ground state.
%
%The interaction between the two may be both repulsive, corresponding to an FF ground state, or attractive,  to an LO ground state.   

%

\section{\label{appendix_mean_field}On the mean-field calculations}
For completeness we recapitulate the variational mean-field formulation in terms of a Hamiltonian $H_{\text{MF}}=\sum_i\mu_iA_i$, where $A_i$ are (normal and anomalous) quadratic fermion operators and $\mu_i$ are variational parameters.\cite{chaikin1995principles} The parameters $\mu_i$ are chosen to minimize the free energy $\Omega=\av{H}_{\text{MF}}-\mu \av{N}_{\text{MF}}-T S_{\text{MF}}$. With $S_{\text{MF}}$ the entropy corresponding to the density matrix of the quadratic Hamiltonian and $H$ the full Hamiltonian. Stationary points $\frac{\partial\Omega}{\partial\mu_i}=0$ correspond to the equations 
\begin{equation}
\sum_j(\frac{\partial\av{H-\mu N}_{\text{MF}}}{\partial\av{A_j}_{\text{MF}}}-\mu_j)\frac{\partial\av{A_j}_{\text{MF}}}{\partial\mu_i}=0 \, ,
\end{equation}
such that the variational parameters should satisfy 
\begin{equation}
\mu_i=\frac{\partial\av{H-\mu N}_{\text{MF}}}{\partial\av{A_i}_{\text{MF}}} \, .
\end{equation}
Importantly, stationarity implies that the full set of quadratic operators $A_i$ generated by a complete Wick decomposition of $\av{H}_{\text{\text{MF}}}$ should be included in the variational Hamiltonian. 

For the problem at hand with the proposed variational parameters $\{\Delta_{\ve{Q}_1}(\ve{k}),\Delta_{\ve{Q}_2}(\ve{k}),\mu_\ve{k}\}$ (where $\mu_\ve{k}$ couples to $\sum_{\sigma}c^\dagger_{\ve{k}\sigma}c_{\ve{k}\sigma}$) this means that we also generate terms related to CDW and higher harmonic pair-fields, as discussed in the main text. (This is true for the LO states where both orders are non-zero. For the FF state, similarly to an ordinary single component superconductor, the mean-field Hamiltonian is complete.) 

Considering the lowest order CDW terms $\rho_{\ve{q}}(\ve{k})=\sum_\sigma\av{c^\dagger_{\ve{k}\sigma}c_{\ve{k}+\ve{q}\sigma}}_{\text{MF}} $ with $\ve{q}=\pm(\ve{Q}_1-\ve{Q_2})$ we can quantify its importance by calculating the energy in the state acquired by solving the approximate equations \eqref{approx_1} and \eqref{approx_2}. Here, the CDW magnitudes are calculated from the expression for the off-diagonal Greens function \eqref{off-d} and \eqref{off-d2}, to lowest (2nd) order in $\Delta$. (I.e. neglecting higher order anomalous Greens functions other than $\scr{F}^*_{\ve{k}-\ve{Q}_{1/2},\ve{k}}$.) The energy related to the CDW order is then given by 
\begin{equation}
 \begin{split}
 &E_{\text{CDW}}=\\
 &-\! \! \frac{V}{N^2}\! \!  \!  \! \! \!  \sum_{\substack{\ve{k},\ve{k}',\\ \ve{q}=\ve{Q}_1,\ve{Q}_2}}\! \! \!  \! \! v(2\ve{k}^{\LCp}) g_{s,d}(\ve{k}^{\LCm}\!\! + \!   \ve{q})g_{s,d}(\ve{k}^{\LCm})  \rho_{\LCm  \ve{q}}(\ve{k})\rho_{ \ve{q}}(\ve{k}')
 \end{split}
 \end{equation}
 where $\ve{k}^{\LCpm}=\frac{\ve{k}\pm \ve{k}'}{2}$, and $\rho_{\ve{q}}(\ve{k})=\frac{1}{\beta}\sum_{n}\scr{G}_{\ve{k},\ve{k}+\ve{q}}(i\omega_n)$. In contrast to the superconducting condensate the Gaussian $v(2 \ve{k}^{\LCp})$ will restrict the summation over $\ve{k}^{\LCp}$. In general we can make $E_\text{CDW}$ negligible by picking a small enough $\kappa_{x,y}$. Indeed $\frac{E_\text{CDW}}{E-E_N}  \lesssim 0.01$ for parameters used in this paper. The energy contribution of the CDW is tiny, consequently, we expect a relatively small CDW order-parameter and that the coupling between different harmonics of the pair-fields are weak such that we can treat these independently for each set of $\ve{Q}_1$ and $\ve{Q}_2$.

An additional aspect of the interaction \eqref{interaction} is that it breaks the discrete rotational symmetry of the tight-binding lattice. Therefore the Hartree-Fock (density-density) terms will generate an anisotropic single-particle energy. (For a nearest-neighbor attraction, the additional terms would be momentum independent in standard fashion and absorbed in $\mu$.) The variational single-particle energy now reads
\begin{equation}
\mu_{\ve{k}}=\varepsilon_{\ve{k}}-\mu-\frac{1}{N}\sum_{\ve{k'}}V(0,0,\ve{k}+\ve{k}')\av{n_{\ve{k}'}}+\text{exchange}\,,
\end{equation}
with $V$ given by \ref{V_form} and where ``exchange'' indicates a similar (slightly more complicated) term arising from the exchange. This implies that the single-particle energies also need to be solved for self-consistently, even in the normal state, leading to a nematic distortion of the Fermi-surface. A numerical check of the magnitude of the contribution of the Hartree-Fock terms to the energy indicates that they are of similar magnitude and sign as the condensation energy. However, for a proper study of this effect it would seem appropriate to also include a (local) Coulomb interaction which would counteract the distortion. Although an interesting topic for future studies we have ignored this aspect in the present work, and taken $\mu_{\ve{k}}=\varepsilon_{\ve{k}}-\mu$.

For completeness, we also comment on the role of the chemical potential $\mu$ in these calculations. We are working with fixed particle number but with different variational Hamiltonians specified by the two pair momenta $\ve{Q}_1$ and $\ve{Q}_2$ (one momentum for the FF states). For each realization, 
the chemical potential must be tuned in order to get the correct particle number, while minimizing the free energy $\Omega$ within the variational space. Clearly, at the correct particle number $N$ the state which minimizes $\Omega$ (corresponding to some value of $\mu$) also minimizes the free energy $F=\av{H}_{\text{MF}}-T S_{\text{MF}}$ at this particle number. It is the latter $F$ that should be used as a measure to compare mean-field solutions at different $\ve{Q}_{1,2}$ when working at fixed particle number. 

\section{\label{appendix_entropy}Entropy}
The excitations of our mean-field Hamiltonian can in principle be found as Bogoliubov quasi-particles $\gamma_{\alpha} = \sum_{i} A_{i \alpha} c_{i} + B_{i \alpha} c^{\dagger}_{i}$ where $i$ range over $2N$ degrees of freedom (spin and momenta) and $\alpha$ over $2N$ quasi-particles. The entropy can then be found through the standard expression
\begin{equation}
S = - \sum_{\alpha} f_{\alpha} \ln f_{\alpha} + (1 - f_{\alpha} ) \ln (1 - f_{\alpha} )
\end{equation}
where $f_\alpha$ is Fermi-Dirac distribution of $\varepsilon_{\alpha}$. Using $\sum_{i} |A_{i \alpha}|^2 + |B_{i \alpha}|^2 =1$ we can make contact with the sum over all single-particle Greens function $\sum_{i}\scr{G}^{\text{singl.}}_i(z) = \sum_{\alpha, i} \frac{|A_{\alpha, i}|^2}{z-\varepsilon_{\alpha}} +  \frac{|B_{\alpha, i}|^2}{z+\varepsilon_{\alpha}}$. Thus the entropy can be written as weighted sum over all residues of the full set of Greens functions $\scr{G}_i$
\begin{equation}
S = - \sum_{p} \text{Res}(p) \left( f_{p} \ln f_{p} + (1 - f_{p} ) \ln (1 - f_{p} ) \right) \, 
\label{entropy}
\end{equation}
where $f_{p}$ is evaluated at the corresponding pole $\varepsilon_p$. Since we truncate the Gorkov equations by throwing away off-diagonal Greens function we do not have access to the exact single-particle states, however, we can still use expression \eqref{entropy} as a consistent approximation of entropy.

\section{\label{appendix_homogenous_current} Depairing current and critical current}

The physics of a supercurrent is treacherous subject with various limits and considerations. Considering a homogeneous current flowing in a superconductor is formally justified in a 3D material in the limit $d \ll \xi, d \ll \lambda$, where $d$ is the sample size and $\xi, \lambda$ the correlation and penetration depth respectively. $d \ll \lambda$ implies that we can neglect the effect of self-field from the current,\cite{tinkham1996introduction,bulaevskii2011vortex} thus we set $\ve{A} = 0$ (given that there is no background field). The limit $d \ll \xi$ ensures that the order-parameter can be taken homogeneous.\cite{likharev1979superconducting} However this regime can be hard to accomplish for cuprates and the depairing limit is not anticipated to be reached because of two main reasons: (i) Inhomogeneous current distribution in samples of sizes larger than the penetration depth\cite{aslamazov1983resistive} $d \gg \lambda$, and (ii) proliferation of vortices\cite{bulaevskii2011vortex} for sample-sizes bigger than the coherence-length $d \gg\xi$. Realizing the limit $d \ll \lambda $ is often not a problem for thin enough samples since the effective penetration depth is governed by the Pearl length\cite{pearl1964current} $\lambda_P = \lambda^2 / d_t$, where $d_t$ is the sample thickness. With $\xi$ typically quite small, the limit $d \ll \xi$ is harder to accomplish.\cite{xu2010achieving} However in the intermediate limit, $\xi \ll d_w \ll \lambda_P, \, d_t \sim \xi$ (with $d_w$ being the sample width) the onset of resistivity due to vortices is only slightly lower than the depairing current,\cite{bulaevskii2011vortex} thus the calculation considered in this paper should hold even in this limit.

In 2D, where the relevance of a PDW has been discussed in connection to the decoupling of layers at $1/8$ doping in LBCO,\cite{li2007two,berg2007dynamical} resistivity occur at any finite current.\cite{minnhagen1987two} Thus the influence of topological excitations on the PDW-current should be considered. Nevertheless, a superconducting order will remain until the depairing current is reached, i.e resistivity reach normal values and the diamagnetic response will disappear. Thus the presented analysis should give an upper estimate to the critical current.

% ========================= BIBLIOGRAPHY ====================

\bibliography{Bib_archive}

\end{document}